\documentclass[twocolumn]{aastex63}
\usepackage{natbib}
\usepackage[utf8]{inputenc}
\usepackage[normalem]{ulem}
\hypersetup{linkcolor=red,citecolor=cyan,filecolor=cyan}

\usepackage{booktabs}

\newcommand{\msun}{\mbox{$M_\odot$}}
\newcommand{\afe}{[$\alpha$/Fe]}
\newcommand{\feh}{[Fe/H]}

\newcommand\edel{\bgroup\markoverwith
{\textcolor{red}{\rule[0.5ex]{2pt}{0.8pt}}}\ULon}

\submitjournal{ApJS}
\shortauthors{Yi \& Jang et al.}
\graphicspath{{./}{figures/}}
\usepackage{lipsum}
\usepackage{amsmath}
\usepackage{comment}
\usepackage{multirow}
\usepackage{enumitem}
\usepackage{threeparttable}
\usepackage{hyperref}
\usepackage{changepage}
\usepackage{amsmath}

\begin{document}
\inputencoding{utf8}
\title{On the Significance of the Thick Disks of Disk Galaxies}

\author[0000-0002-4556-2619]{Sukyoung K. Yi}
\affil{Department of Astronomy and Yonsei University Observatory, Yonsei University, Seoul 03722, Republic of Korea}

\author[0000-0002-0858-5264]{J. K. Jang}
\altaffiliation{Email address: starbrown816@yonsei.ac.kr}
\affiliation{Department of Astronomy and Yonsei University Observatory, Yonsei University, Seoul 03722, Republic of Korea}

\author[0000-0002-8140-0422]{Julien Devriendt}
\affil{Department of Physics, University of Oxford, Keble Road, Oxford OX1 3RH, UK}

\author[0000-0003-0225-6387]{Yohan Dubois}
\affil{Institut d’Astrophysique de Paris, Sorbonne Université, CNRS, UMR 7095, 98 bis bd Arago, 75014 Paris, France}

\author[0000-0001-9939-713X]{San Han}
\affiliation{Department of Astronomy and Yonsei University Observatory, Yonsei University, Seoul 03722, Republic of Korea}

\author[0000-0002-3950-3997]{Taysun Kimm}
\affil{Department of Astronomy and Yonsei University Observatory, Yonsei University, Seoul 03722, Republic of Korea}

\author[0000-0001-6180-0245]{Katarina Kraljic}
\affil{Universit\'e de Strasbourg, CNRS, Observatoire astronomique de Strasbourg, UMR 7550, F-67000 Strasbourg, France}

\author[0000-0002-8435-9402]{Minjung Park}
\affiliation{Center for Astrophysics $\vert$ Harvard \& Smithsonian, 60 Garden St., Cambridge, MA 02138, USA}

\author{Sebastien Peirani}
\affil{Observatoire de la Côte d’Azur, CNRS, Laboratoire Lagrange, Bd de l’Observatoire,Université Côte d’Azur, CS 34229, 06304 Nice Cedex 4, France}
\affiliation{Institut d’Astrophysique de Paris, Sorbonne Université, CNRS, UMR 7095, 98 bis bd Arago, 75014 Paris, France}

\author[0000-0003-0695-6735]{Christophe Pichon}
\affiliation{Institut d’Astrophysique de Paris, Sorbonne Université, CNRS, UMR 7095, 98 bis bd Arago, 75014 Paris, France}

\author[0000-0002-0184-9589]{Jinsu Rhee}
\affiliation{Korea Astronomy and Space Science Institute, 776, Daedeokdae-ro, Yuseong-gu, Daejeon 34055, Republic of Korea}
\affiliation{Department of Astronomy and Yonsei University Observatory, Yonsei University, Seoul 03722, Republic of Korea}

\inputencoding{utf8}

\begin{abstract}
Thick disks are a prevalent feature observed in numerous disk galaxies including our own Milky Way. 
Their significance has been reported to vary widely, ranging from a few to 100\% of the disk mass, depending on the galaxy and the measurement method. 
We use the NewHorizon simulation which has high spatial and stellar mass resolutions to investigate the issue of thick disk mass fraction. 
We also use the NewHorizon2 simulation that was run on the same initial conditions but additionally traced nine chemical elements. 
Based on a sample of 27 massive disk galaxies with $M_* > 10^{10}$\,\msun\ in NewHorizon, the contribution of the thick disk was found to be $34 \pm 15$\% in $r$-band luminosity or $48 \pm 13$\% in mass to the overall galactic disk, which seems in agreement with observational data.
The vertical profiles of 0, 22, and 5 galaxies are best fitted by 1, 2, or 3 $\mathrm{sech}^2$ components, respectively. 
The NewHorizon2 data show that the selection of thick disk stars based on a single \afe\ cut is severely contaminated by stars of different kinematic properties while missing a bulk of kinematically thick disk stars. 
Vertical luminosity profile fits recover the key properties of thick disks reasonably well.
The majority of stars are born near the galactic mid-plane with high circularity and get heated with time via fluctuation in the force field.
Depending on the star formation and merger histories, galaxies may naturally develop thick disks with significantly different properties.
\end{abstract}

\section[]{Introduction}
\label{intro}

Well over half of the massive galaxies are said to be ``spiral'' in morphology, but what is perhaps more striking than the beautiful spiral patterns is their thin disk. 
Most of the massive spiral galaxies are thin disks in essence. 
The vertical scale height of the Milky Way (MW) disk is only about 300 pc \citep{Juric2008TheDistribution}, while the radial scale length of the thin disk is an order of magnitude larger \citep{Bland-Hawthorn2016TheProperties}.

The thin nature of disk galaxies is often attributed to the assembly history of angular momentum from the neighborhood during the disk development: tidal torque theory \citep{Peebles1969OriginGalaxies}.
In addition, a coherent assembly of angular momentum and secular diffusion of the galaxy's orbital structure over time results in a disk galaxy \citep{Binney1988,Pichon2011,Halle2018}. 
In this view, the morphology of a galaxy is heavily influenced by the ``cosmic web".

A thicker component of the galactic disk was found in the Milky Way \citep{Gilmore1983} and other galaxies \citep[e.g.,][]{Burstein1979,Yoachim2008TheGalaxies,Comeron2018}. 
The initial star count studies conducted on the Milky Way disk found that the vertical distribution of disk stars showed a break and thus was better fitted by two exponential components rather than one \citep[e.g.,][]{Gilmore1983}. 
The mass fraction of the ``thick disk" was initially derived to be only a couple of percent, and its vertical scale height was measured to be several times that of the thin disk.
The MW thick disk was later fitted with $\mathrm{sech}^2$ function \citep{vanderKruit1981}, and the fraction of the thick disk stars was estimated to be about 4\% \citep{Juric2008TheDistribution}.
Dynamical modeling applied to the star count data considering the orbital motions of stars embedded in a dark matter halo suggested 14\% for the thick disk mass contribution \citep{Bovy2013AKpc}, which makes the thick disk no longer such a negligible component as was initially hinted.
An explanation is required for such prominent structures.

A spectroscopic effort for finding thick disk stars was made in parallel. 
A flurry of observations has noted that stars in the Solar Neighborhood are bimodal in \afe\ abundance \citep[e.g.,][]{Bensby2006, Reddy2006, Lee2011FormationSample}. 
A naive projection of the \afe\ bimodality onto kinematic properties leads to a suggestion that the thick disk fraction is roughly a third within 1 kpc from the galactic mid-plane \citep{Lee2011FormationSample}.
This makes the thick disk more significant than previously believed. 
Does the \afe\ bimodality indicate a thin-to-thick disk mass ratio?
Why does it suggest a different mass fraction for the thick disk from vertical profile fit studies?

Thick disks are found in other galaxies, too. 
\citet{Yoachim2006StructuralGalaxies} inspected 34 edge-on disk galaxies with a wide range of mass.
They measured the vertical scale heights and luminosity fractions of their thick disks. 
The vertical scale height ratios between the thick and thin disks were found to be 2--3.
The luminosity fraction of the thick disk ranges widely between 0 and 100 \%,
and the Milky Way values are compatible with these data.

The origin of thick disks has been a subject of debate.
The stars in spiral galaxies were probably born predominantly in the molecular clouds within thin disks.
Such a pre-existing disk of stars would be kinematically heated through various sources of perturbation such as minor mergers \citep{Quinn1993HeatingMergers, Kazantzidis2008ColdAccretion}, spiral arms and bars \citep{SellwoodJ.A.1984SpiralFormation,Roskar2013TheDiscs,Vera-Ciro2014heating}, or giant molecular clouds \citep{Spitzer1951TheVelocities}. 
The accreted stars from satellite galaxies are likely to be kinematically hotter occupying a thicker region of the disk \citep{Abadi2003SimulationsDisks}. 
``Resonant thickening'' of the disk by small satellites has also been suggested \citep{Binney1988,sellwood1998}.
Accretion of gas clouds may induce star formation away from the galactic disk mid-plane and thus occupying a thicker region of the disk \citep{Norris1991}.
Alternatively, gas-rich galaxy mergers could result in active star formation away from the thin disk. 
The stars formed this way would have a higher probability of occupying a thicker region of the disk \citep{Brook2004TheUniverse, Agertz2021}.

It is clear that the issue of thick disks is outstanding. 
How significant is it?
Why is its significance in the Milky Way estimated to differ according to different methods?
What are the main mechanisms behind it?
Can we understand the range of thick disk contributions in other galaxies?
These are the questions we want to answer through this investigation.

To answer these questions, we use the NewHorizon (NH) simulation \citep{Dubois2021}.
We also use the NewHorizon2 simulation (NH2, Yi et al. in prep) which is a lower-resolution twin that started from the same initial conditions but has a more extensive chemical follow-up (see Section 2).
The combination of the high resolution of NH and detailed chemical information of NH2 makes this investigation viable.

\begin{figure*}
\centering
\includegraphics[width=0.85\textwidth]{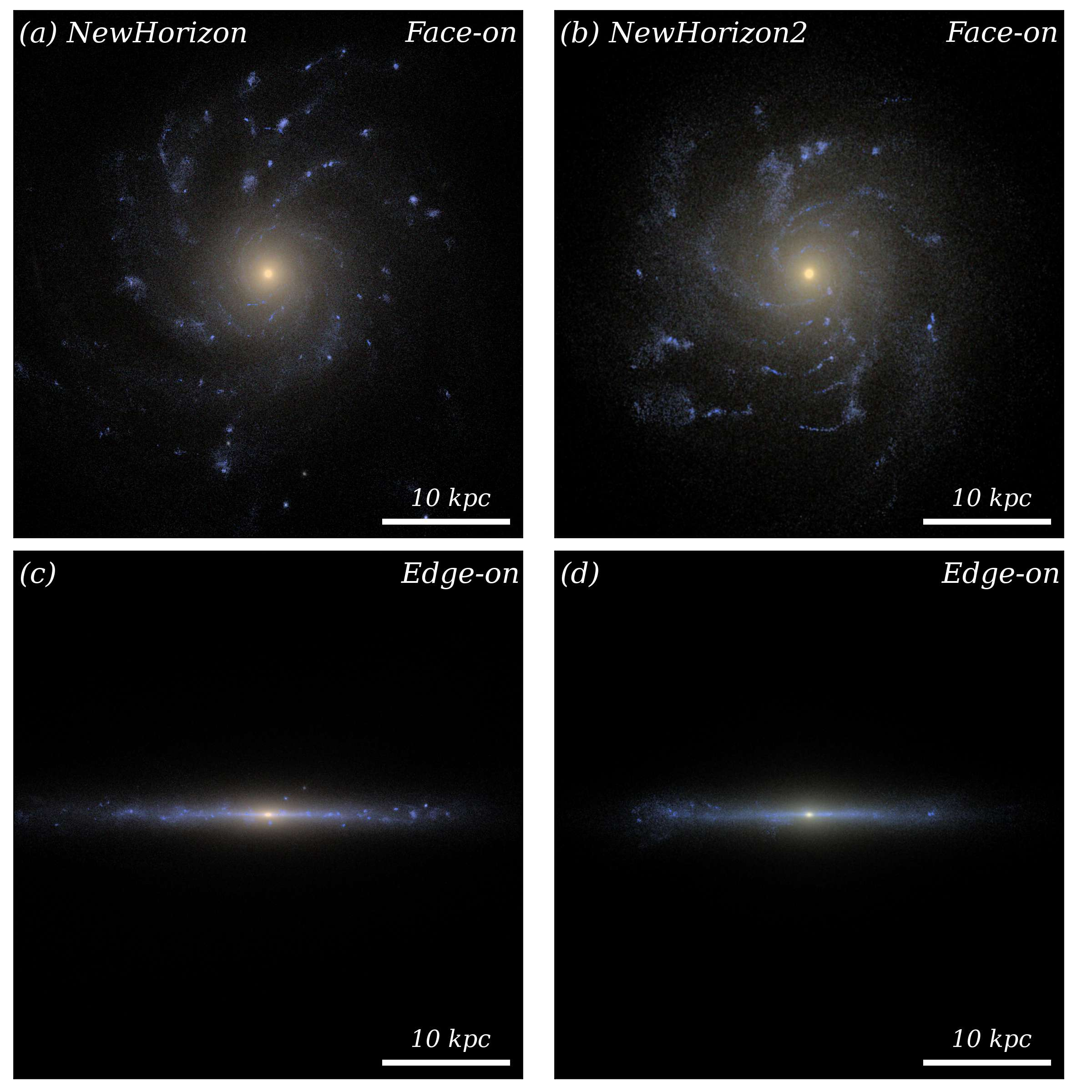}
\caption{
The face-on and edge-on images of sample galaxies from NH (Panels a, c, also in Appendix Figure~\ref{fig_appendixA1} as NH ID: 8) and from NH2 (Panels b, d, also in Appendix Figure~\ref{fig_appendixA2} as NH2 ID: 2). 
Red, green, and blue colors correspond to SDSS $i$-, $r$-, and $g$-band fluxes, respectively.
}
\label{fig_images}
\end{figure*}

\section{Data}
\label{data}

The main data used was the NH simulation \citep[][]{Dubois2021}. 
NH is a medium-sized (20 Mpc diameter), cosmological hydrodynamic simulation of galaxy formation which was run with RAMSES \citep{Teyssier2002CosmologicalRefinement}.
It resolves dense regions spatially down to 34 pc (at $z=0$) which allows for sampling the vertical structure of a Milky Way-size disk in 10 bins or more, which is ideal for tackling the issues of the galactic disk structure.
Its stellar mass resolution is $10^4$~\msun.
Therefore, typical Milky Way-mass galaxies are realized by millions of star particles.
The temporal resolution is also exceptionally high: 867 snapshots were stored with a mean timestep of 15 Myr.
This allows a precise inspection of the motions of the star particles and gas.
The NH volume contains 48 galaxies of stellar mass above $10^{10}$~\msun\, and so provides good statistics for drawing a general picture on thick disks.
Because of its relatively small volume, only some limited comparison against observations is possible. 
It has been shown that NH reproduces some of the key properties of galaxies: the stellar mass function of galaxies, surface brightness-stellar mass relation, star formation rate density evolution, Kennicutt-Schmidt relation, galaxy mass-size relation, and mass-metallicity relation to name a few. 
Readers are referred to \citet{Dubois2021} for a detailed description of NH.

We also use the NH2 simulation.
It starts from the same initial conditions as NH but has run with a factor of two coarser spatial resolutions (the best resolution of 68 pc) allowing for quicker execution than NH.
There are other minor differences between NH and NH2. 
For example, NH2 has a 50\% higher stellar feedback efficiency in the hope of matching the empirical stellar to halo mass relation better.
The most significant advantage of NH2 is that it traces nine chemical elements (H, D, C, N, O, Mg, Si, S, and Fe) and provides useful chemical information for star formation history studies.
This allows us to enhance the understanding we earlier had on thick discs based on the NH simulation alone \citep{Park2021}. 
In this investigation, we use the NH data for most analyses.
However, when the chemical information is inspected, we use the NH2 galaxies. 
We do not mix NH and NH2 for the same galaxy because, although NH and NH2 start from the same initial conditions, a few processes adopt a random choice, e.g., star formation \citep{Rasera2006,Kimm2017Feedback-regulatedReionisation}, and thus cause a  difference in the final properties of the galaxy. 

For the chemical evolution of galaxies, we take into account the contributions from stellar winds, as well as SN Ia and II.
To calculate the yields from a simple stellar population (SSP) for each stellar particle, we use the Starburst99 model \citep[][]{Leitherer1999}, assuming a Chabrier initial mass function \citep[][]{Chabrier05}.
The Geneva stellar wind model \citep[][]{Schaller92, Maeder00} is used to calculate the tabulated chemical yields from the stellar wind as a function of the age and metallicity of the SSP (see \citealp{Leitherer14} for more details).
Therefore, each stellar particle is involved in chemical enrichment based on its evolving age and metallicity.
SN II explosions from 8--50\,\msun\ stars are responsible for the chemical elements returned by each SSP.
The SN II yields are based on \citet{Kobayashi06}, where the yields from intermediate-mass stars (8--13\,\msun) are taken from \cite{Woosley95}.
The chemical yields of SN Ia are obtained from \citet{Iwamoto99}.
To estimate the SN Ia frequency for each stellar particle, we adopt the delay time distribution model described in \citet{Maoz12}.
This model assumes a power-law decline of the SN Ia frequency over time ($\propto t^{-1}$), which allows us to calculate the SN Ia event rate for a given age of the stellar particle.
Based on the normalization values provided in \cite{Maoz12}, each stellar mass particle of $10^4$~\msun\ is expected to have at most 13 SN Ia explosions during its lifetime.

Our galaxy mass threshold ($M \geq 10^{10}$~\msun) allows at least one million star particles in each galaxy.
Of the 48 NH galaxies, we remove 3 galaxies that are severely disturbed morphologically at $z=0.17$.
We finally have 27 disk galaxies of $M_* \geq 10^{10}$~\msun\ with $v/\sigma \geq 1$ for stars in the last snapshot of $z = 0.17$, where $v/\sigma$ stands for the ratio between the rotation speed and velocity dispersion.
Similarly, we have selected 10 massive disk galaxies from the NH2 simulation. 
We have fewer galaxies in the final NH2 sample than in NH primarily for two reasons. First, the factor of two coarser spatial resolution reduced $v/\sigma$.
Besides, its enhanced stellar feedback caused the final stellar mass to be smaller in NH2.
We present the images of the sample galaxies from NH and NH2 in Figure~\ref{fig_images} and Figures~\ref{fig_appendixA1}, \ref{fig_appendixA2} in Appendix~\ref{sec appendix_A}.

\section{Results}

\subsection{Profile Sampling}
\label{profile}

With its 34 pc ``best'' spatial resolution of NH in the adaptive mesh refinement scheme, we effectively resolve the disks of massive galaxies. 
In addition, its mass resolution of $10^4$~\msun\ for star particles is also essential for dynamically realizing the galactic disk in numerical simulations.
Only when we could reproduce the thin disk in the first place would our attempt to understand the thick disk be justified. 

\begin{figure}
\centering
\includegraphics[width=0.35\textwidth]{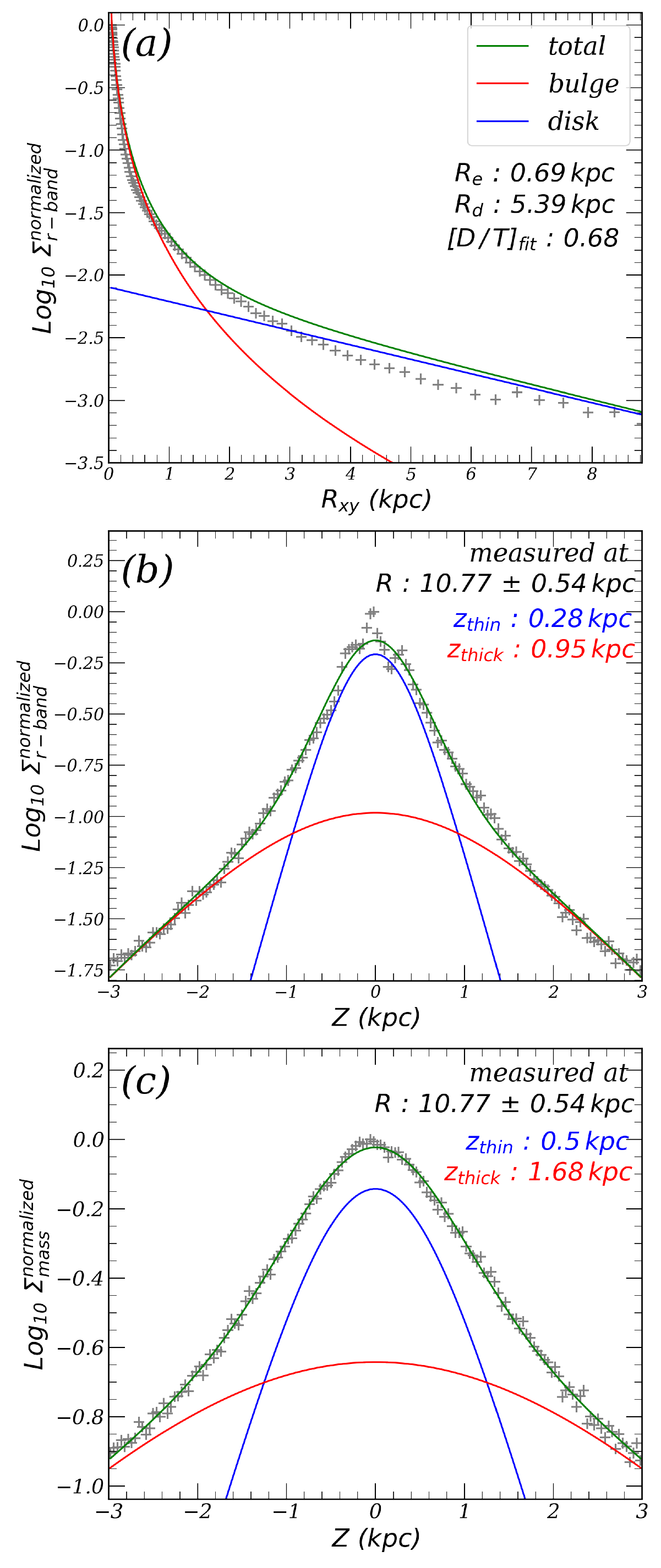}
\caption{
The two-component fit to the profiles of the galaxy shown in Figure 1 (NH ID: 8). The plus symbols show the actual profiles and the continuous lines show the fits.
(a) The radial profile of the galaxy (face on) is well reproduced by a combination of two components: an exponential disk and a spheroid with a Sersic index 4.  
(b) The vertical {\em luminosity} profile at $(2.0 \pm 0.1) R_{\rm d}$ and $|z| \leq 3$~kpc is well reproduced by a two-component model. 
(c) The vertical {\em mass} profile appears slightly different from the mass profile but is still closely reproduced by a two-component model. 
}
\label{fig_profile}
\end{figure}

To measure the structural properties of a disk, we need to choose a location in the disk that is representative of the disk. 
We first measure the scale length of the disk, $R_{\rm d}$, using all the star particles inside $R_{\rm 90}$, where $R_{\rm 90}$ stands for the radius within which 90\% of the stellar mass resides.
We found a linear relation between $R_{\rm d}$ and the galaxy's stellar mass, $M_*$. 
We use $R_{\rm d}$ based on this fit, and we measure the thin and thick disk contributions at the radial distance from the galaxy center $2.0 \pm 0.1 R_{\rm d}$ and the vertical distance from the mid-plane $|z| \leq 3$~kpc using the $\mathrm{sech}^2$ function \citep[e.g.,][]{Yoachim2008TheGalaxies, Comeron2011ThickBaryons, Park2021}. 

\begin{figure*}
\centering
\includegraphics[width=0.9\textwidth]{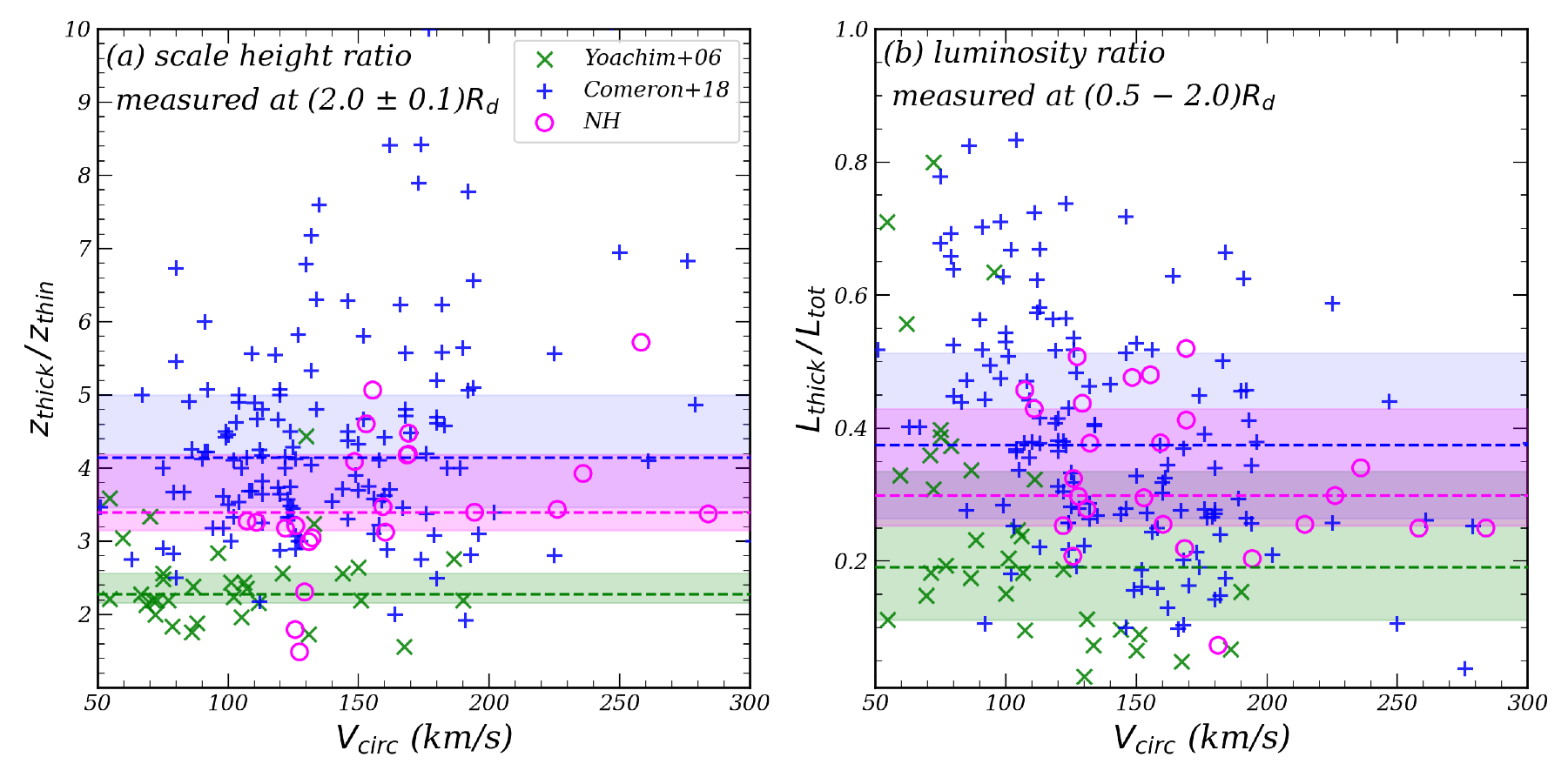}
\caption{
The measured properties of the thin and thick disks of the NH galaxies in r-band compared with the observational data.  
Circles denote the NH galaxies.
(a) The ratios in the vertical scale height between thick and thin disks. 
(b) The luminosity fraction of the thick disks. 
The dashed lines and shades with a matching colors show the median and first-third quantiles of the three data sets.
}
\label{fig_observation}
\end{figure*}

We attempt to fit the vertical mass and luminosity profiles of NH galaxies by allowing up to 3 components: 
\begin{equation}
    \rho\,(\rm Z ) = \sum_{i=1}^3 \rho_{i}\,\rm sech^2\!\left( \frac{|Z|}{2\,{\it z}_{i}} \right)
\end{equation}
where $i=1$, 2, 3 are for the ``thin'', ``thick'', and ``thicker'' disk components, and where $\rho_{\rm i}$ and $z_{\rm i}$ are the density and vertical scale height of each component. 
We use the Bayesian information criteria (BIC) to choose the best-fit model, where the BIC is a measurement that uses both the log-likelihood of the fit (the goodness of the fit) and the number of parameters. 
As we use more parameters in statistical analysis, the BIC value will be penalized more.
Based on the BIC analysis, we found that 0, 22, and 5 galaxies are best fitted by 1, 2, or 3 vertical components, respectively.
We will discuss this later in Section~\ref{conclusion}. 
We provide the BIC-selected fits for a few massive NH galaxies in Appendix (Figure~\ref{fig_appendixA3} \& Figure~\ref{fig_appendixA4}).

Figure~\ref{fig_profile} shows the sample two-component fits to the profiles of the sample galaxy presented in Figure~\ref{fig_images}. 
Panel-(a) shows the radial surface brightness profile fit to the face-on images using the Sersic profiles \citep{Sersic1963InfluenceGalaxy}.
A simple combination of a disk with a Sersic index $n=1$ and a spheroid with $n=4$ reproduces the radial profile reasonably. 
The two lower panels show the vertical profile fits in mass and luminosity using the $\mathrm{sech}^2$ function.
While they appear somewhat different from each other, the mass and luminosity profiles are equally well-fitted by a model with two components that could be called the ``thin'' and ``thick'' disks.
The disk properties of this model galaxy seem compatible with those of the observed galaxies (see below).

We now restrict ourselves to a two-component scenario to make our results directly comparable to previous studies which were predominantly based on the assumption of two disk components. 
In other words, we present the two-component fits to the vertical profiles of the NH galaxies. 
In the case of galaxies for which 3-component fits were selected as the best by the BIC, the mass fractions of the third components are very small.

Figure~\ref{fig_observation} shows the vertical scale height ratios between the thin and thick disk components and the luminosity fractions of the thick disks of the NH galaxies compared with the observed data \citep{Yoachim2006StructuralGalaxies, Comeron2018, Martinez-Lombilla2019}. 
This figure is an updated version of Figure~4 of \citet{Park2021}. 
It is based on the galaxies at $z=0.17$ and on the measurements at a new position on the disk mentioned above.
The thick disk properties of the NH galaxies are in reasonable agreement with the observed data, particularly those of \citet{Comeron2018}.
There is a hint of a negative mass ($V_{\rm circ}$) dependence in the luminosity fraction, in Panel-(b), when the observed data of \citet{Comeron2018} and the NH galaxies are considered, where the circular velocity $V_{\rm circ}$ has been measured from the stellar particles inside $R_{\rm 90}$ in case of the NH simulation.

The thick disk luminosity fraction was measured by integrating over the range of 0.5--$2.0 R_{\rm d}$.
The mean value of the integrated luminosity fractions of the thick disks of the NH galaxies is $34 \pm 10$\%, where the errors are the standard deviation among galaxies.
When we fit the mass profiles as in the bottom panel of Figure~\ref{fig_profile}, the contribution of the thick disks becomes $54 \pm 10$\% instead.
This indicates that the thick disks play a significant role in the overall mass distribution of the galaxies.

In summary, the NH simulation resolves the thin disks of spiral galaxies. 
The vertical distribution of stars indicates the presence of  thicker disk components in all of our sample galaxies.
When we consider a two-component assumption consisting of thin and thick disks, the properties of the thick disks in the NH galaxies show good agreement with the observed data.
Thick disks contribute roughly one third of the disk in $r$-band luminosity or half in mass.
Thick disks are not minor.

\subsection{Spatial Sampling}
\label{spatial}

\begin{figure}
\centering
\includegraphics[width=0.45\textwidth]{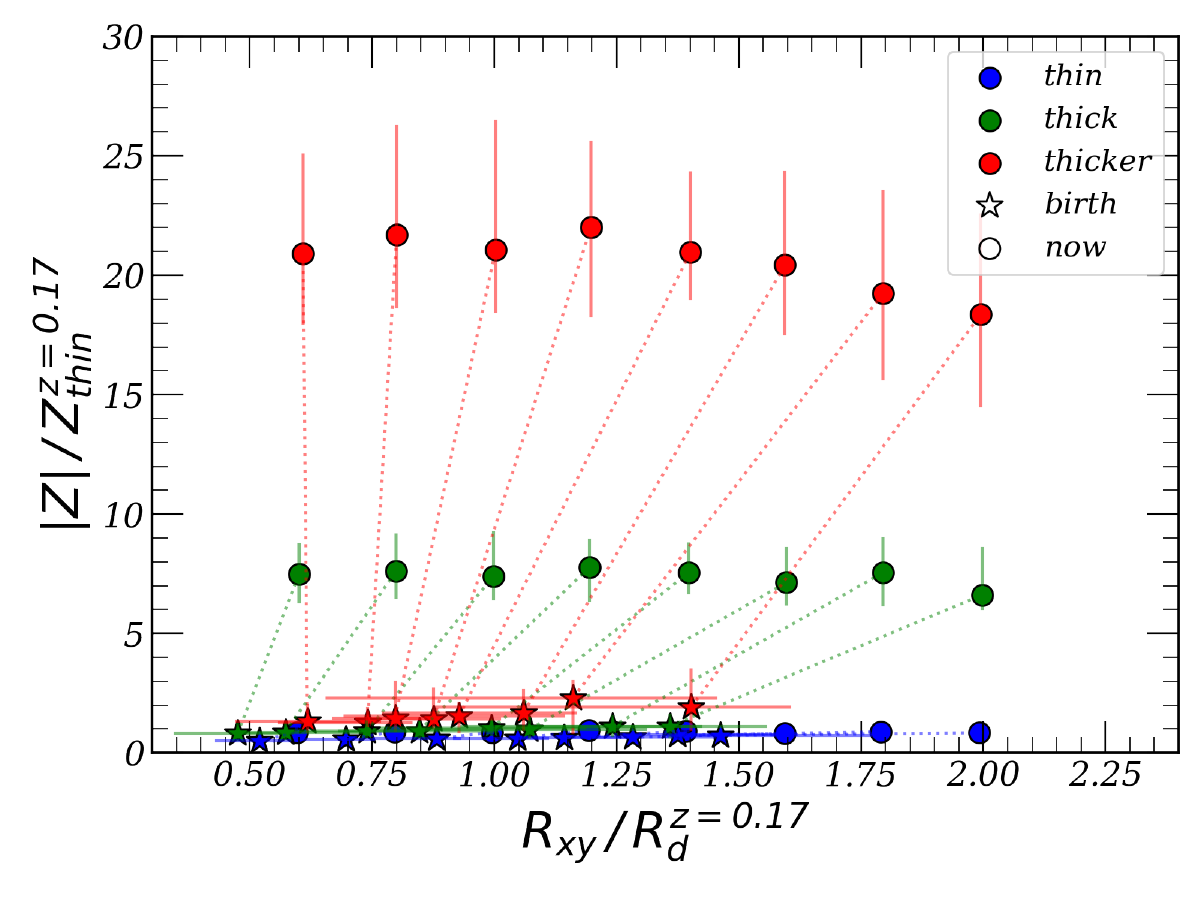}
\caption{
The positions of the spatially-thin and -thick disk stars of all the 27 NH galaxies when they are observed (final snapshot) and at their birth. The abscissa shows the mean radial position of the stars from the galactic centre in the units of the disk scale length in the final snapshot. The ordinate shows the mean position of the stars in the vertical distance from the galaxy disk's mid-plane in units of the thin disk scale height in the final snapshot. The three spatially separated groups of stars in the final snapshot (circles) were barely distinguishable in terms of birth position (star symbols). 
The thin ($|z|/z_{\rm thick} < 0.5$; blue), thick ($1.5 < |z|/z_{\rm thick} < 2.5$; green), and thicker ($5 < |z|/z_{\rm thick} < 6$; red) component's median positions are shown with the first and third quantiles, where $z_{\rm thick}$ is roughly 0.8~kpc.
}
\label{fig_birthplace}
\end{figure}

We would like to study the properties of thin and thick disk stars and their origins.
The Solar Neighborhood is composed of both thin and thick disk stars, and it is not trivial to know which star belongs to which component.
One way of evading this difficulty, at least in part, is to sample the stars spatially.
Based on the definition of thin and thick disks, it is natural to assume that the fraction of thick disk stars decreases with the vertical distance from the disk's mid-plane.
We fit the vertical $r$-band brightness profiles of the NH galaxies using two $\mathrm{sech}^2$ components in the cylindrical region at $2 \pm 0.1 R_{\rm d}$.
We consider the stars in the mid-plane at vertical distance $|z|/z_{\rm thick} \leq 0.5$ ``spatially-thin'' disk stars because they are dominated by the thin-disk component according to the two-component fit.
We consider the stars away from the mid-plane
``spatially-thick'' disk stars based on the same arguments.
We select two locations to sample thick disk stars: $1.5 < |z|/z_{\rm thick} < 2.5$ and $5 < |z|/z_{\rm thick} < 6$, where $z_{\rm thick} \approx 1$~kpc.
An advantage of this sampling method is that it can be easily applied to observed edge-on galaxies.
\citet{Park2021} showed from their sample of 18 galaxies that there is little difference in terms of the birth position between the spatially-thin and -thick disk stars. 
We present in Figure~\ref{fig_birthplace} the birth positions of the spatially-thin and -thick disk stars of the 27 NH galaxies and confirm their earlier report.
Both axes are normalized to the scale values measured in each radial bin in the final snapshot.
Both spatially-thin and -thick disk stars are born near the galactic mid-plane within 1--2 vertical scale heights of the thin disk in the final snapshot ($|z| < z_{\rm thin}^{z=0.17}$), where the mean value of $z_{\rm thin}^{z=0.17}$ is roughly 0.2~kpc.

Spatial sampling is useful for learning the properties of thin and thick disk stars, provided that the contamination is not severe. 
However, it does not provide us with the mass contribution of thick disk stars.

\begin{figure}
\centering
\includegraphics[width=0.37\textwidth]{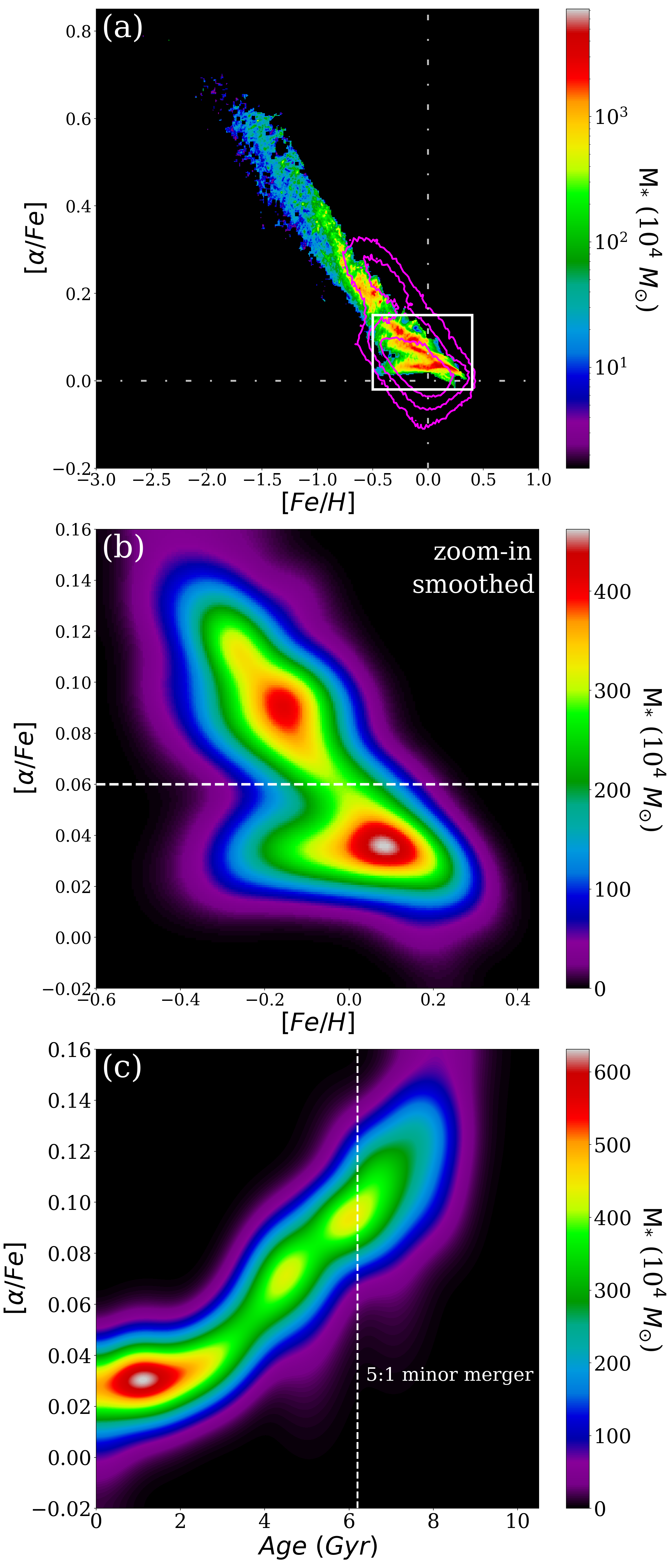}
\caption{The stellar population properties of the star particles of a sample NH2 galaxy  (NH2 ID: 2). (a) The density map of all the star particles in terms of chemical properties. The cyan contours show the Milky Way stellar data adopted from the APOGEE \citep{Jonsson2020_APOGEE_DR16} and GALAH \citep{Buder2021_GALAH_DR3} projects. The dash-dotted lines show the zero points, and the white box shows the region that is zoomed in Panel-(b). (b) The zoom-in map of the box in Panel-(a). This smoothed distribution appears to be bimodal which is similar to what is found in the Milky Way disk. One may divide the bimodal distribution at the valley marked by the horizontal dashed line at [$\alpha$/Fe]=0.06, which in turn classifies the stars into thin- (below the line) and thick-disk (above the line) stars. (c) \afe\ vs. age for the star particles in Panel-(b). The bimodality in [$\alpha$/Fe] is clearly associated with the age distribution and hence the star formation history. The coalescence time for the gas-rich minor merger at 6.2 Gyr of lookback time, discussed in the main text, is marked.} 
\label{fig_chemical}
\end{figure}

\subsection{Chemical Sampling}
\label{chemical}
As we mentioned in Section~\ref{intro}, the mass fraction of the Milky Way thick disk is estimated to be 33\% when \afe\ is used as the sole classifier, while star counts suggest much lower values.
We would like to check what the simulation models suggest.
In the NH2 simulation, we monitor the chemical evolution of galaxies in multiple elements including alpha elements (e.g., O, and Mg).
We derive \afe\ based on the formula: $\rm [\alpha/Fe] = [(O + Mg + Si)/3Fe]$, where the solar abundance, $\rm [O/H]_\odot$, $\rm [Mg/H]_\odot$, $\rm [Si/H]_\odot$, and $\rm [Fe/H]_\odot$ are adopted from \citet{Asplund2009TheSun}. 
We present the density distribution of the chemical properties of the star particles of one sample galaxy in Figure~\ref{fig_chemical}. 
Panel-(a) shows the [$\alpha$/Fe] vs. [Fe/H] plane. 
Overall, there is a reverse correlation between these properties. 
This is naturally expected for a largely ``closed-box'' system, considering the chemical yields of SN Ia and II \citep{Tinsley1980}.
The majority of stars have high values of [Fe/H], and as can be seen later, they are the stars that belong to the disk. 
The highest density peak has $\rm [Fe/H] \approx 0.1$ and $\rm [\alpha/Fe] \approx 0.03$ which roughly agree with the values of the disk stars of the Milky Way \citep[e.g.,][]{Bensby2014}.

\begin{figure}
\centering
\includegraphics[width=0.4\textwidth]{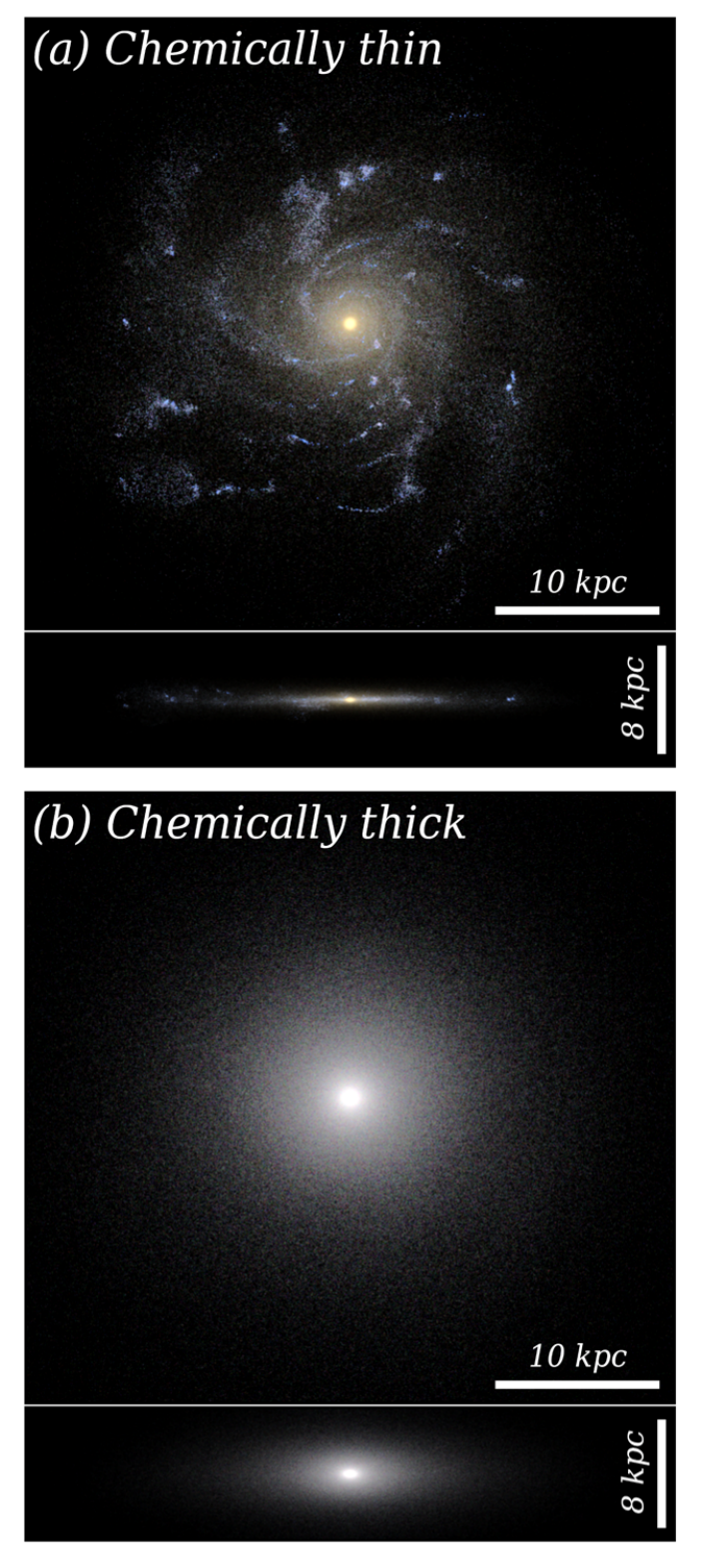}
\caption{
The color images of the thin (top row) and thick-disk (bottom row) stars classified by \afe\, as illustrated in Figure~\ref{fig_chemical}-(b)  (NH2 ID: 2). Seemingly the \afe=0.06 cut is effective as a thin-thick disk classifier.
} 
\label{fig_morph_chem}
\end{figure}

Figure~\ref{fig_chemical}-(b) shows a zoom-in map of a region of Panel-(a). 
The smoothed map clearly shows a bimodal distribution.
A simple cut of \afe=0.06 effectively separates the bimodal distribution, and the mass fraction of the ``chemically-thick'' ($0.06 \leq [\alpha/\rm Fe] < 0.14$) component becomes 34\% when we measure it at $2 R_{\rm d}$ and $|z| \leq 1$ kpc for easy comparison with the observational estimate. 
In the case of Milky Way, bimodality appears to be separable by a cut of $[\alpha/\rm Fe] \approx 0.2$, and the number ratio between the high-\afe\ peak and the low-\afe\ peak stars is 1:2 \citep{Lee2011FormationSample}, leading to a thick-disk mass fraction of 33\%.

The apparent bimodality of the Milky Way disk in alpha abundance is often attributed to two separate populations of disk stars and even to two distinct origins. 
Figure~\ref{fig_chemical}-(c) shows the age distribution of the same stars shown in Panel-(b). 
It is evident that the alpha abundance distribution is directly linked to the age distribution and hence the star formation history. 
In the case of this particular galaxy, there was a  minor merger with a mass ratio of 5:1 around 6.2\,Gyr ago (in coalescence time), which caused a starburst with rough properties of $\rm [Fe/H] \approx -0.2$, and $\rm [\alpha/Fe] \approx 0.09$.
If we consider this model galaxy as an analog of the Milky Way, the bimodality of the alpha abundance found in the Milky Way disk could be interpreted as a result of a galaxy merger that happened in the past. 
This explanation was already suggested by \citet{Agertz2021} based on the Vintergatan simulation. 
However, the reality is more complex.
We will discuss in the next section why a high alpha peak cannot be interpreted as the result of a merger event alone.

The case of the NH2 galaxy shown in Figure~\ref{fig_chemical} has a bimodal distribution similar to that of the Milky Way, but the exact value that divides the bimodality is different: Milky Way's cut appears to occur at \afe=0.2 while that of the NH2 galaxy does at 0.06. 
The detailed shape of the bimodality (or multimodality) is specific to the unique star formation history of the galaxy in question.
We provide the \afe-[Fe/H]-age maps (just Panels b and c) for the NH2 galaxies in Appendix (Figures~\ref{fig_appendixA5}, \ref{fig_appendixA6}).
One can see that galaxies show a wide variety of the \afe\ distribution.

Figure~\ref{fig_morph_chem} shows the stellar density distribution of the two components based on a cut of $\rm [\alpha/Fe]=0.06$.
For consistency in terms of nomenclature, we call the stars with $\rm [\alpha/Fe] \leq 0.06$ ``chemically-thin'' and $\rm [\alpha/Fe] > 0.06$ ``chemically-thick''.
Panel-(a) shows the chemically-thin disk stars and Panel-(b) shows the chemically-thick disk stars.
Chemically-thick disk stars appear to occupy a thicker region of the disk compared to chemically-thin disk stars.
Cursory inspection of this figure suggests that \afe\ is effective for finding thin and thick-disk star candidates.
In the next section, we further inspect this finding in connection with kinematic information.
The fraction of chemically-thick disk stars near the Solar Circle is 34\% in mass (or 8\% in luminosity) in this galaxy, compared to 33\% in the MW according to the SDSS observations.
The proximity between the two values was most likely achieved by chance.

\subsection{Kinematic Sampling}
\label{kinematic}

An important advantage of using simulation data is that we have detailed kinematic information for the galaxies. 
Thanks to the high mass, spatial, and temporal resolutions of NH, each galaxy is resolved by millions of star particles with detailed orbital information.
Recent studies \citep{Du2019,Du2020,Jang2023} have utilized the Gaussian Mixture Model (GMM), which is an unsupervised machine learning technique that finds a given number of groups or kinematic components in the kinematic phase space. 
It uses three quantities: total energy $e$, angular momentum 
along the rotational axis $J_{\rm z}$, and the remaining angular momentum $J_{\rm p}$, where $J_{\rm p} = J_{\rm tot}-J_{\rm z}$.

\begin{figure}
\centering
\includegraphics[width=0.43\textwidth]{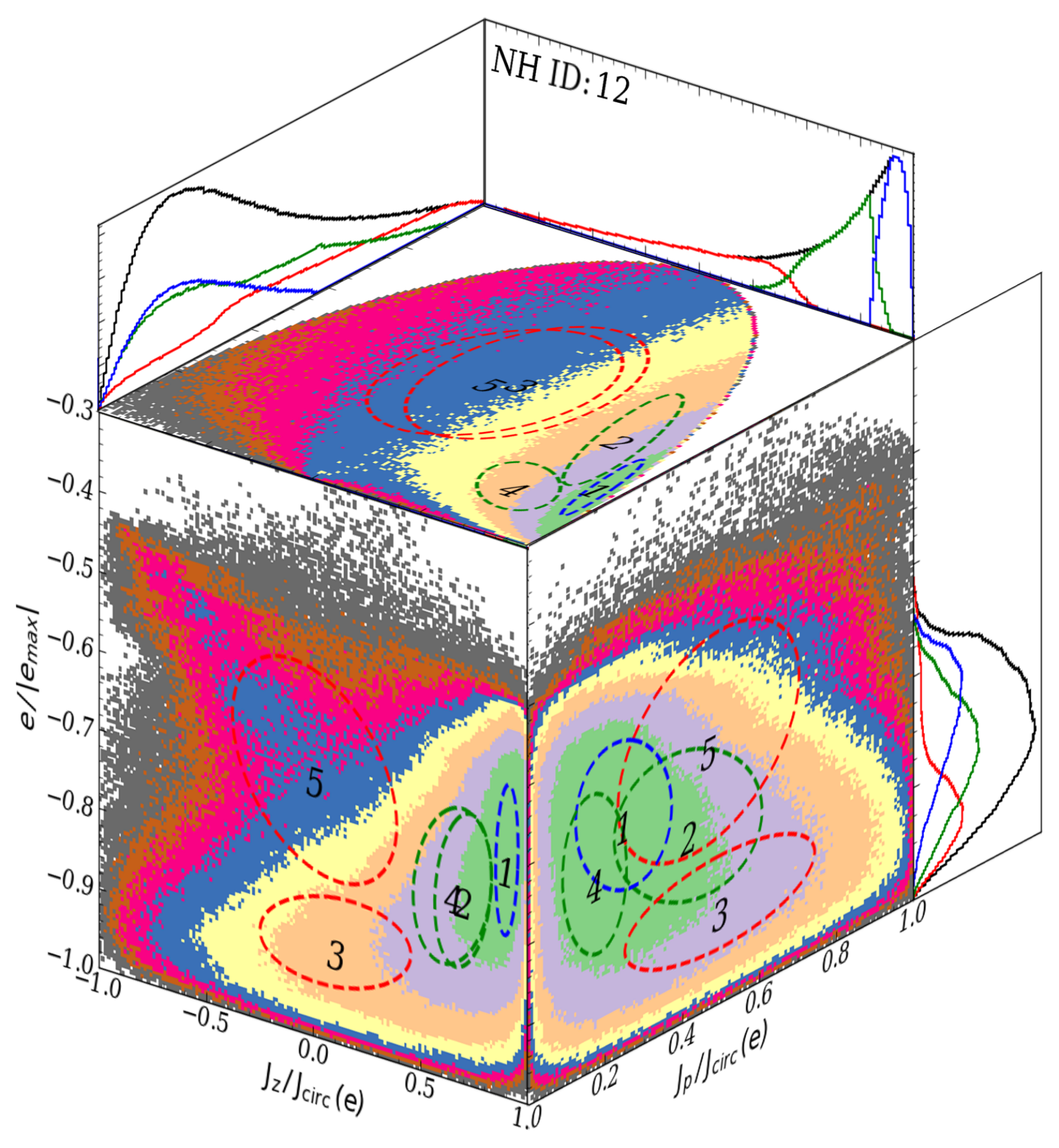}
\caption{
The kinematic properties of all the star particles of a sample NH galaxy. 
This GMM exercise identified the five components in order of mass fraction based on three properties ($e/|e_{\rm max}|$, $J_{\rm z}/J_{\rm circ}(e)$, and $J_{\rm p}/J_{\rm circ}(e)$).
The dashed ellipses show the thin disk (Component 1 marked in blue), thick disk (Components 2 and 4 in green), and spheroid (Components 3 and 5 in red) components.
The projected histograms associated with the three groups (blue, green, and red curves) along with the total (black) are also presented.
} 
\label{fig_gmm}
\end{figure}

Figure~\ref{fig_gmm} shows a sample exercise of GMM on the galaxy in Figure~\ref{fig_images}.
The diagram shows the energy $e$ normalized by the maximum energy, and $J_{\rm z}$ normalized by the maximum angular momentum that a particle can have for a specific binding energy $e$, $J_{\rm circ}(e)$. 
The former roughly indicates the mean distance from the galaxy center; the smaller, the closer.
The latter is also known as circularity ($\epsilon \equiv J_{\rm z} / J_{\rm circ}(e)$), where 1 indicates a perfectly-circular rotation with respect to the rotational axis, $-1$ denotes a perfectly-counter-rotating motion, and 0 represents dispersion dominant motion. 
We also note that circularity departure from 1 does not always indicate {\em vertical heating} because the radial migration of a star confined to a razor-thin disk may change its circularity without increasing its vertical displacement. 
However, we found that such cases are rare in our simulations.

\begin{figure}
\centering
\includegraphics[width=0.45\textwidth]{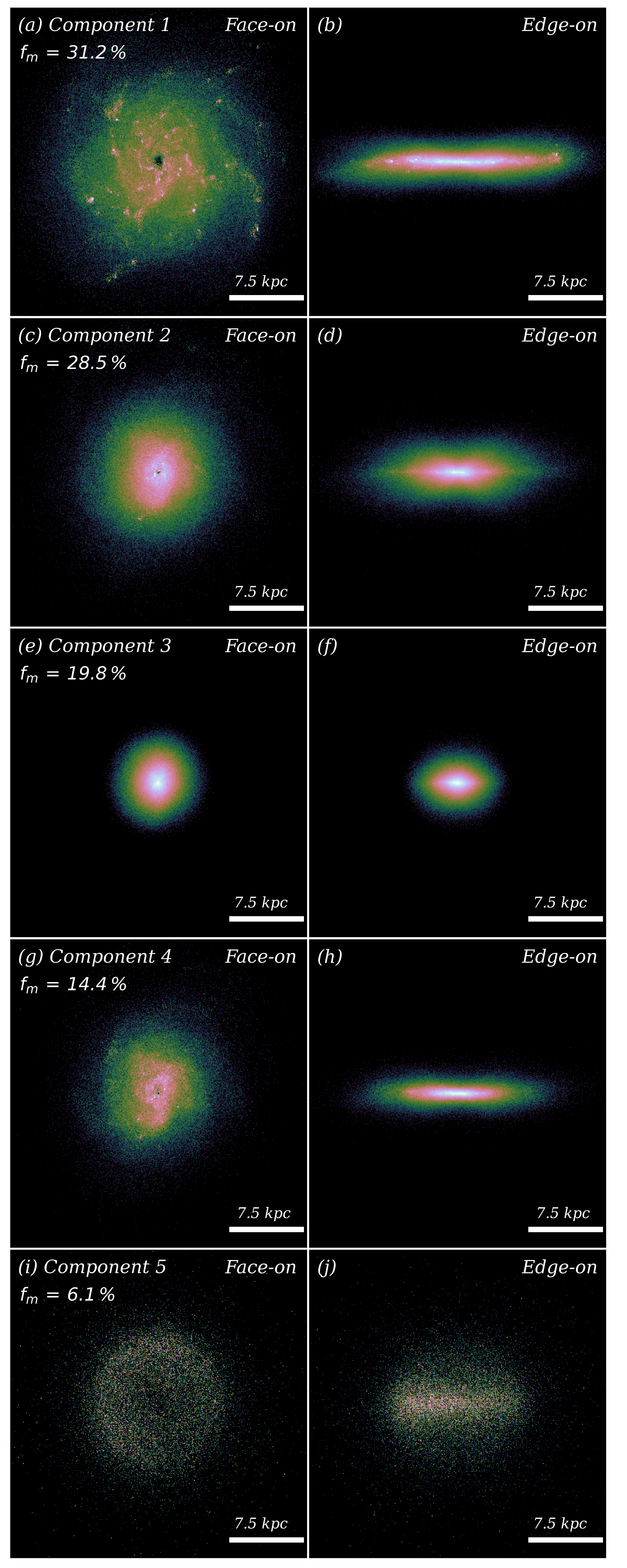}
\caption{
The face-on and edge-on images of the components in Figure~\ref{fig_gmm}.
The mass fraction of each component is shown in each panel.
} 
\label{fig_gmm_spatial}
\end{figure}

In this exercise, we identify the five most significant components by mass contribution, and the numerals in the diagram show the mass rank. 
Components 1, 2, and 4 have $\bar{\epsilon} > 0.5$, where $\bar{\epsilon}$ denotes the mean value of $\epsilon$ of all the members of each component detected.
Therefore, they follow the coherent rotational motion. 
The other components (3 and 5) are centered around $\bar{\epsilon} \approx 0$, meaning that they do not follow the coherent rotation of the galaxy. 
By using a cut in circularity, one may assign the components to the thin disk ($\bar{\epsilon} > 0.85$) or the thick disk $0.5 < \bar{\epsilon} \leq 0.85$.
In this segmentation, the thin disc is therefore defined to be mostly made of quasi-circular in-plane orbits, while the thick component corresponds to orbits with higher vertical or radial velocity dispersion. 
Figure~\ref{fig_gmm_spatial} shows their color maps which clearly have distinct spatial distributions. 
Components 1, 2, and 4 have disky distributions, whereas the other components have dispersed distributions. 
Kinematically-identified components appear to recover the morphology of separate components closely. 
While we detect five components in this exercise, it is possible to explore an arbitrary number of components, in which case even minor features such as the stellar halo and tidal debris could be detected.
We consider the information derived this way ``ground truth'' in this investigation, but it should be noted that the classification of thin or thick disks based on a single cut in circularity is somewhat arbitrary regardless of whether it is through GMM or in any other method.
For reference, we provide the GMM detection maps of additional NH galaxies in Appendix (Figure~\ref{fig_appendixA7} and Figure~\ref{fig_appendixA8}).

\begin{figure}
\centering
\includegraphics[width=0.45\textwidth]{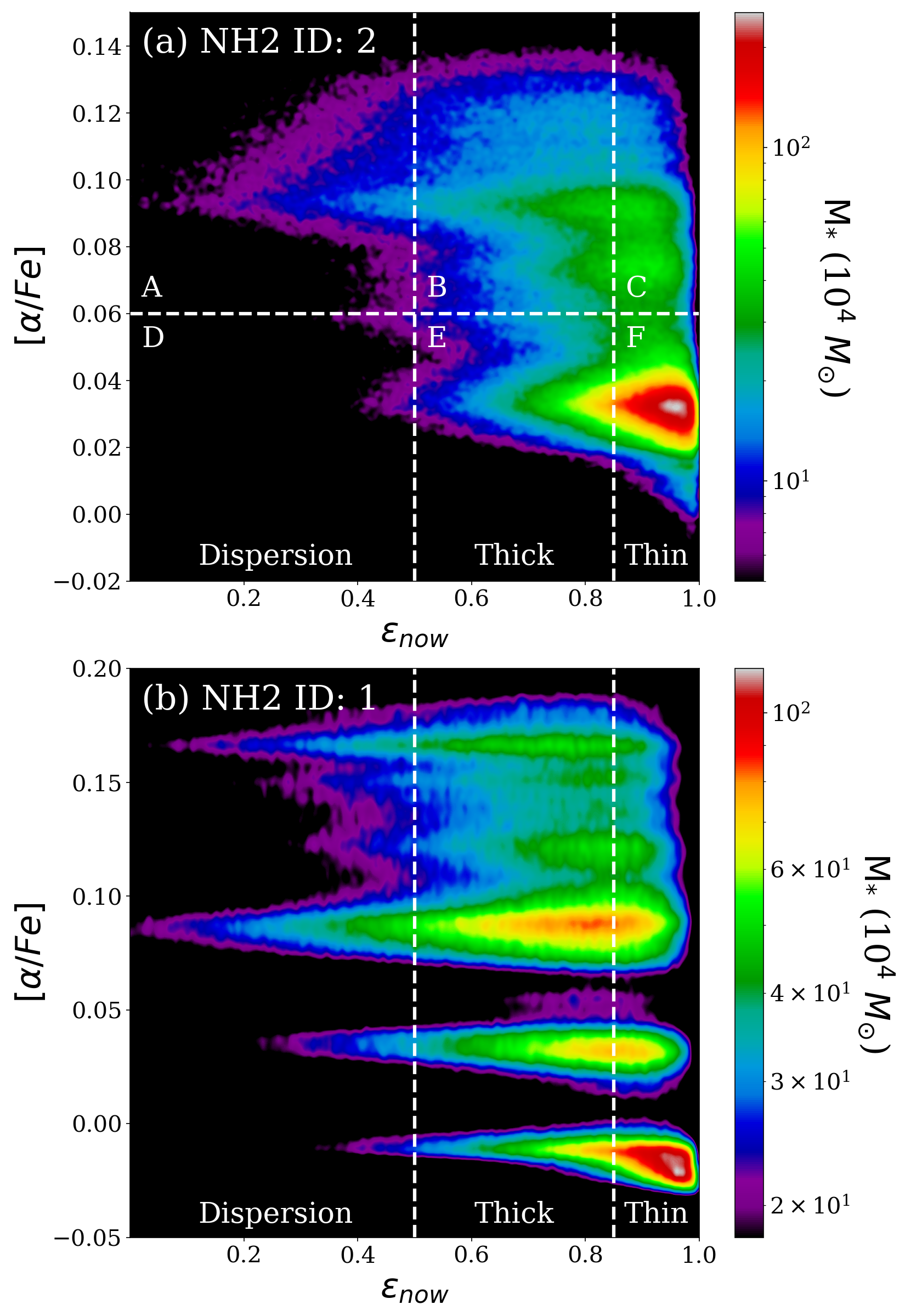}
\caption{
The chemical and kinematic properties of the star particles of two NH2 galaxies. (a) The same galaxy as shown in Figures 5 and 6. The \afe\ bimodality shown in Figure~\ref{fig_chemical} is also visible here. The horizontal dashed line shows the arbitrary cut that separates the low- and high-\afe\ stars from the bimodality. The vertical line at $\epsilon =0.5$ divides rotation-dominant ($\epsilon\geq0.5$) and dispersion-dominant ($\epsilon < 0.5$) stars. The other vertical line at $\epsilon = 0.85$ divides kinematically-thin and -thick disk components. (b) The same information but for a different galaxy. This galaxy has more than three main \afe\ components in the disk instead of two.
} 
\label{fig_alpha_circ}
\end{figure}

Figure~\ref{fig_alpha_circ} shows the star particles in the plane of alpha abundance and circularity.
The top panel shows the same galaxy shown in Figures~\ref{fig_chemical} \& \ref{fig_morph_chem}.
As shown in Figure~\ref{fig_chemical}, a bimodal distribution is visible in \afe. 
A criterion of \afe = 0.06 roughly separates the bimodality in this galaxy.
The low \afe\ clump is dominantly composed of stars with high circularity. 
Therefore, such a simple cut based on \afe\ would effectively select ``kinematically-thin'' disk stars (e.g., $\epsilon > 0.85$). 
However, the situation for the thick disk selection is not so straightforward.
Chemically-thick disk stars (\afe $>0.06$) are not dominated by ``kinematically-thick'' disk stars (e.g., $0.5 < \epsilon \leq 0.85$) at all.
Panel-(a) shows the stars virtually in the whole galaxy at $2 \pm 0.5 R_{\rm d}$ and $|z| \leq 15$~kpc.
Kinematically-thick disk stars (i.e., B=24.9\%) take up only  47.6\% of the chemically-thick disk stars.
The mass fraction of the chemically-thick disk stars (A+B+C) is 52.4\%, while that of the kinematically-thick disk stars (B+E) is 41.0\%. 
If we measure these values near the Solar Circle (e.g., at $2 \pm 0.5 R_{\rm d}$ and $|z| \leq 1$~kpc), they become 35.0\% (chemically thick) and 25.3\% (kinematically thick).
The use of \afe\ as a thick disk indicator substantially overestimates the thick disk fraction.

Figure~\ref{fig_alpha_circ}-(b) shows the case of another galaxy. 
While the galaxy in Panel-(a) shows two peaks in \afe, the one in Panel-(b) shows three or even more peaks.
While the lowest-$\alpha$ peak corresponds to the young thin disk, the other peaks are the results of older stellar populations.
The two peaks at $[\alpha/\rm Fe] \approx 0.04$ and 0.08 are associated with unusually-high star formation events that occurred around 4 and 6 Gyr ago, respectively.
The thick disk of this galaxy has significantly different properties from the galaxy in Panel-(a), and likewise, a variety of thick disk properties are possible in different galaxies.
This illustrates nicely how chemistry allows us to trace back bursts of star formation which then diffuse away in orbital space. 
Stellar chemical information is the time capsule of Galactic archaeology.
We provide \afe\ vs. circularity diagrams for additional NH2 galaxies in Appendix (Figure~\ref{fig_appendixA9}).

\begin{figure*}
\centering
\includegraphics[width=0.9\textwidth]{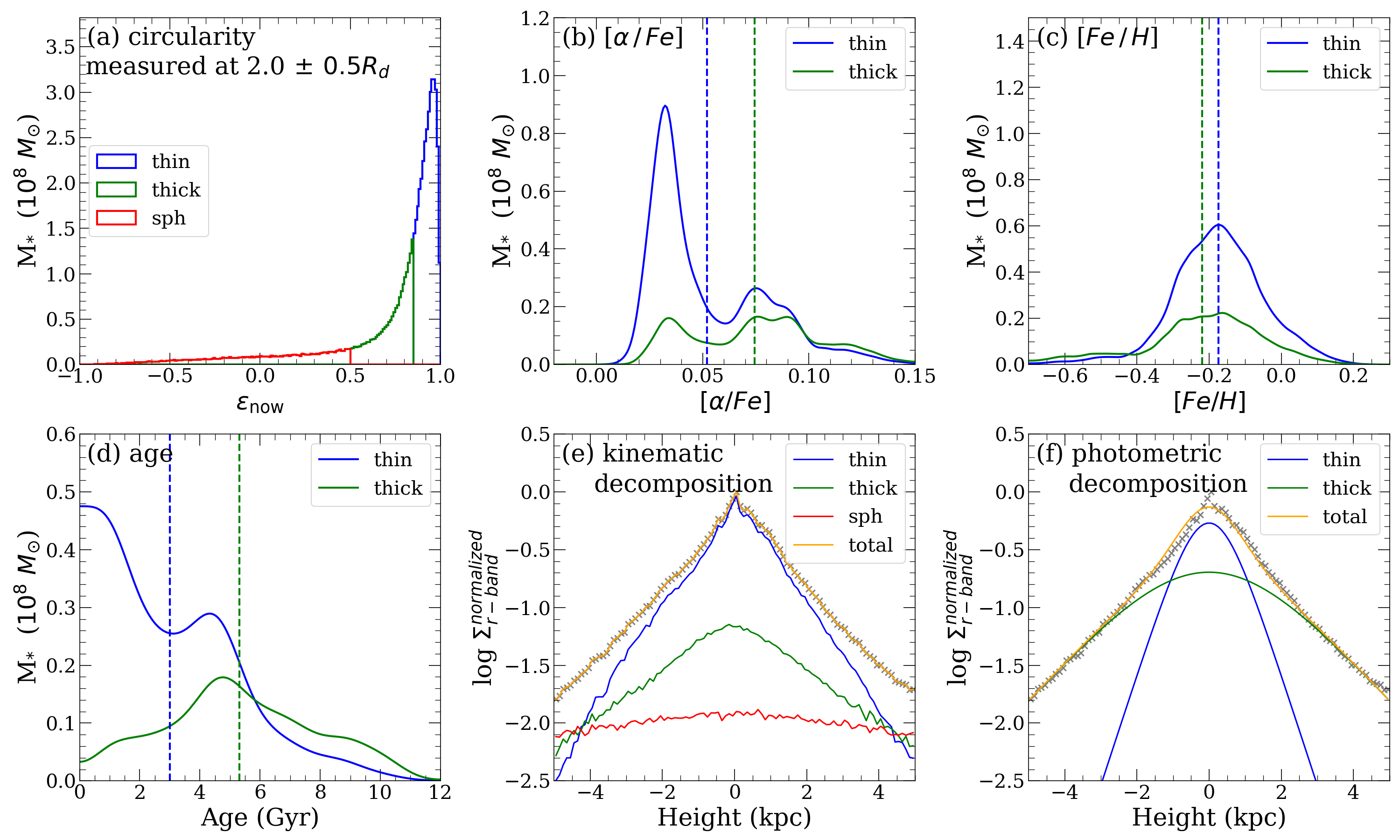}
\caption{
The correspondence between different sampling results based on the NH2 galaxy shown in Figure~\ref{fig_alpha_circ}-(a) (NH2 ID: 2).
(a) A simple circularity-based kinematic sampling scheme. (b) The \afe\ distribution of the thin and thick-disk stars sampled by circularity as shown in (a). 
The two vertical lines show the mean values of the two disk components. As shown in Figure~\ref{fig_alpha_circ}, contamination is severe in the case of the thick disk. (c) Same as Panel-(b) but for \feh. 
(d) Same as Panel-(b) but for age.
(e) The vertical distribution of the thin and thick disk stars sampled in (a). (f) The vertical $r$-band luminosity profile fitted by a two-component model as described in Section~\ref{spatial}.
} 
\label{fig_correspondence}
\end{figure*}

\begin{figure}
\centering
\includegraphics[width=0.4\textwidth]{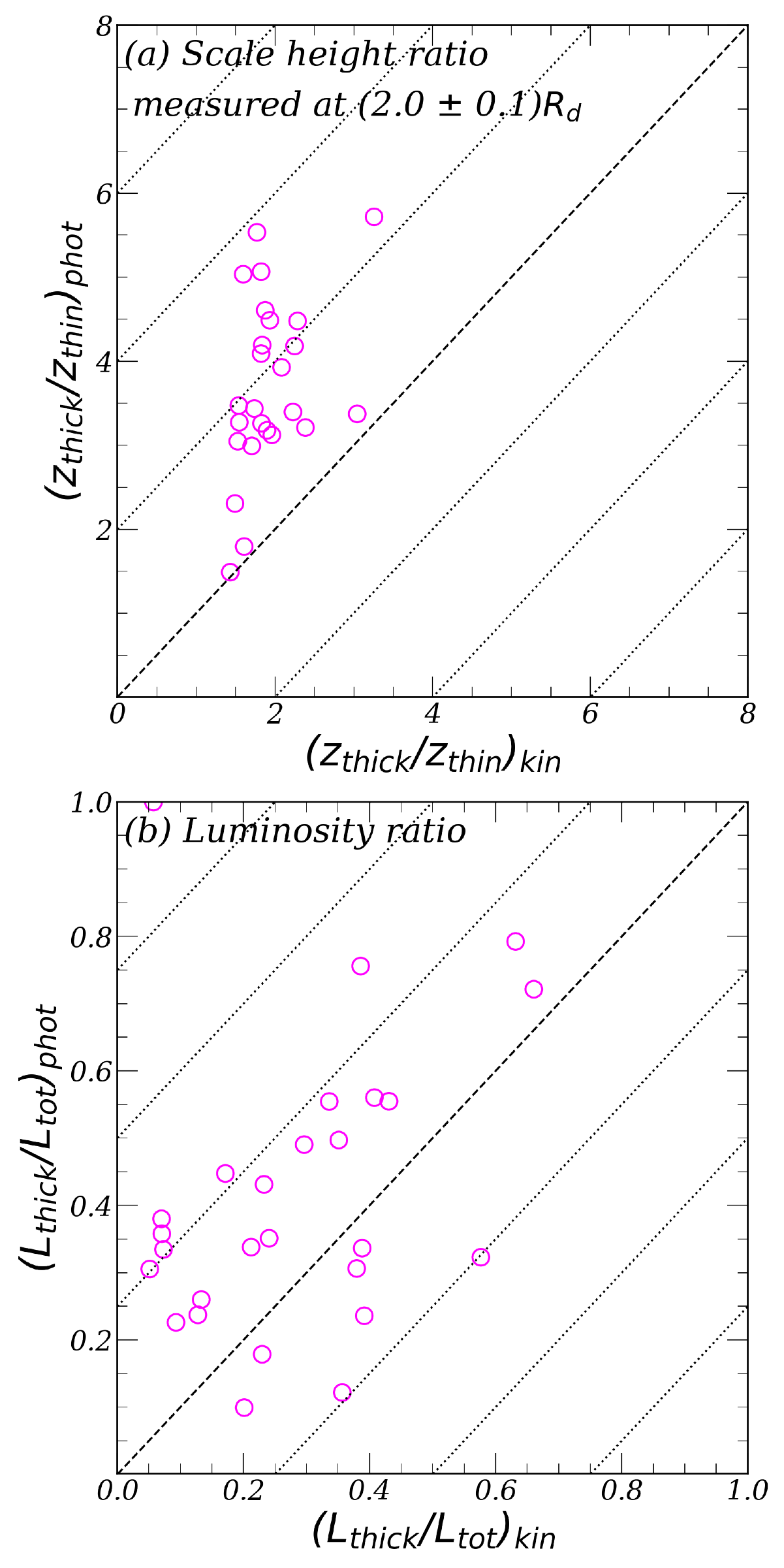}
\caption{
The key properties of thick disks of NH galaxies compared between the photometric vertical profile fit and the kinematic decomposition at $2 R_{\rm d}$. (a) The ratio of the vertical scale heights between the thin and thick disks. (b) The $r$-band luminosity contribution of the thick disks to the total disk luminosity.
} 
\label{fig_photo_kine}
\end{figure}

We check the correspondence between the results from different sampling methods. 
Figure~\ref{fig_correspondence} shows a comparison for the Solar Circle region ($2 \pm 0.5 R_{\rm d}$ and $|z| < 15$~kpc) of a sample NH2 galaxy. 
We here adopt a simple criterion based on circularity instead of using the more elaborate GMM technique: $0.5 < \epsilon \leq 0.85$ for thick disk stars as illustrated in Panel-(a).
The \afe\ distribution based on this criterion is shown in Panel-(b).
As shown in Figure~\ref{fig_morph_chem}, the low-\afe\ peak traces kinematically-thin disk stars well, whereas the high-\afe\ peak is heavily contaminated by kinematically-thin disk stars.
Kinematically-thick disk stars have various values of \afe\ because they are an ensemble of stars born at different times and are thus heated for different lengths of time.
Panel-(c) shows their metallicity distributions.
Panel-(d) shows the age distributions. 
The kinematically-thin disk is predominantly young, whereas the thick disk is composed of stars with a variety of ages.
Their vertical distribution is shown in Panel-(e).
In this galaxy, the kinematically-thin disk is dominant all the way out to 4~kpc from the mid-plane; thus one has to go beyond 4~kpc to sample thick-disk stars.
This is why we presented such distant stars from the galactic plane (red symbols) in Figure~\ref{fig_birthplace}.
Panel-(f) shows a two-component vertical profile fit for comparison.
The thick disk fraction is substantially different from that in Panel-(e) in this galaxy, which explains the large scatter in the luminosity fraction of the thick disk stars (Figure~\ref{fig_photo_kine}).
This highlights the issue regarding the validity of the profile fit and in return the spatial sampling discussed in Section~\ref{profile} and Section~\ref{spatial}.

Figure~\ref{fig_photo_kine} shows the comparison of the key properties of thick disks between the vertical luminosity profile fits and the kinematic detection based on the $0.5 < \epsilon \leq 0.85$ criterion.
Panel-(a) shows the vertical scale height ratios measured at $2R_{\rm d}$.
The profile fits give roughly 50\% higher values.
Panel-(b) shows the thick disk luminosity fraction.
The agreement is somewhat better; the thick disk luminosity fraction is $28 \pm 17$\% (or $43 \pm 14$\% in mass).
When we integrate over 0.5--2~$R_{\rm d}$, the thick disk fraction becomes $34 \pm 15$\% in $r$-band luminosity and $48 \pm 13$\% in mass.
The scatter is large, partly due to the fact that the galaxies are in different dynamical stages.
Besides, the classification of the thick disk depends on the selection criterion, while there is no clear ground that divides thin and thick disks.
The origin of the offset between the two estimates is not obvious to guess at the moment except that we are tempted to say that the profile fits do not recover the true kinematic properties fully.

\begin{figure}
\centering
\includegraphics[width=0.45\textwidth]{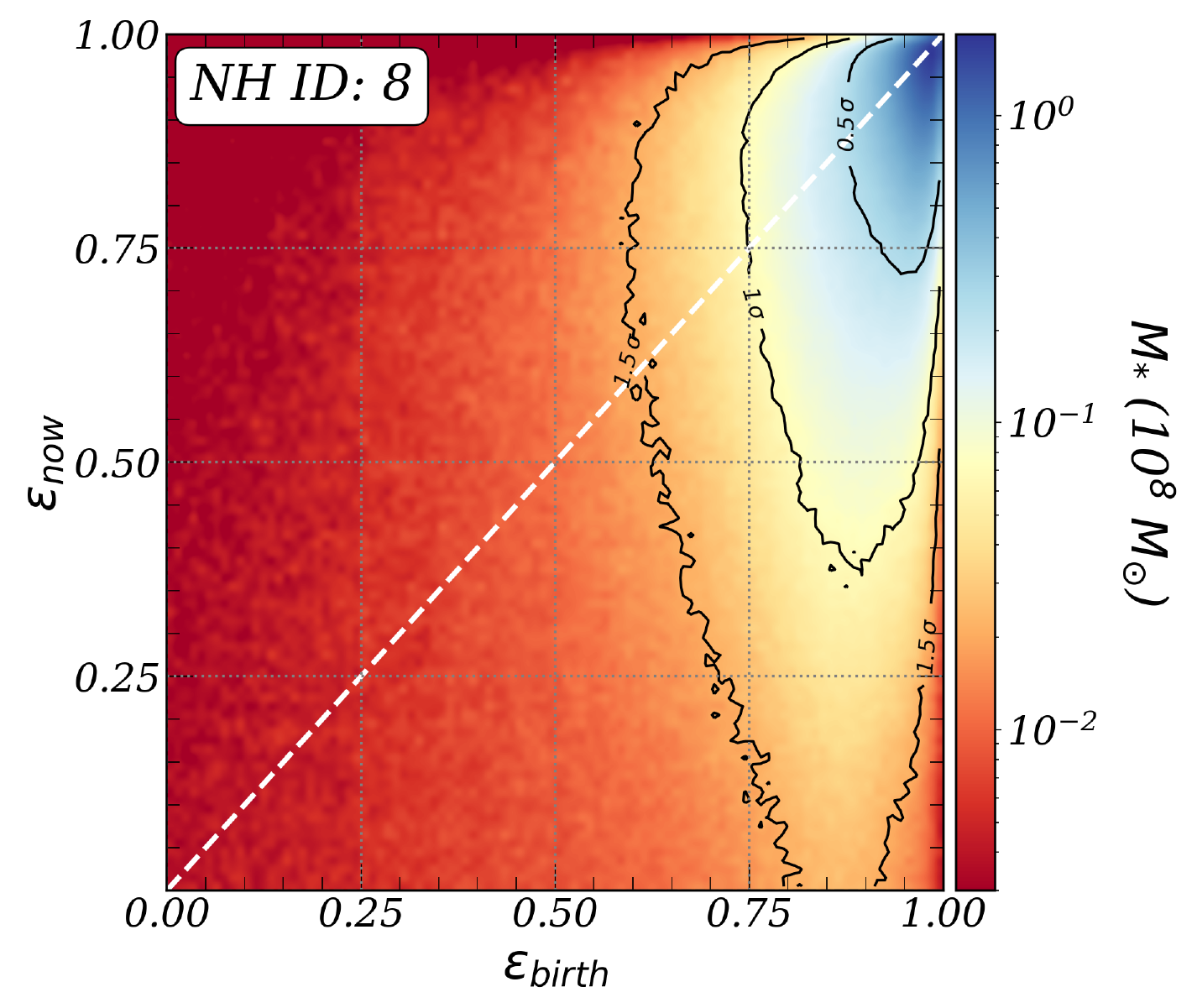}
\caption{
The comparison between the circularity at birth and the circularity today for all the stars of one NH galaxy.
A one-to-one correlation is shown by the diagonal line.
The black curves show the 0.5, 1.0, and 1.5-$\sigma$ contours.
The distribution is heavily biased. Most stars are born with a large value of circularity on a thin rotating disk and are gradually heated to show a much wider distribution in the end. 
} 
\label{fig_circ_birthnow}
\end{figure}

Figure~\ref{fig_circ_birthnow} shows the circularity of all stars in a galaxy when they were born and at present. 
Stars are typically born in highly circular orbits. 
The 1$\sigma$ contour shows that 68\% of the stars are born with $\epsilon \gtrsim 0.75$.
This is because star formation in  typical massive disk galaxies occurs predominantly within thin gas disks.
The circularity distribution of the same stars is much more dispersed at present (i.e., $z=0.17$).
The 1$\sigma$ contour extends down to $\epsilon \approx 0.4$.
This summarizes the general pattern of kinematical heating which corresponds to a slow diffusion of the orbital structure driven by force fluctuation along the unperturbed trajectories. 
In orbital space, this diffusion can in principle operate along three directions, radial migration or ``churning" (change of guiding center), ``blurring" (increase in radial eccentricity), and  ``heating" (increase in the vertical oscillation of stars).
In Figure~\ref{fig_circ_birthnow} we look for simplicity at one projection of these, focussing on a decrease in circularity.
This has also been demonstrated in Figure~\ref{fig_birthplace} based on the spatial/geometric sampling of thin and thick-disk stars.
We provide the circularity at present and at birth for additional NH galaxies in Appendix (Figure~\ref{fig_appendixA10}).

\begin{figure}
\centering
\includegraphics[width=0.35\textwidth]{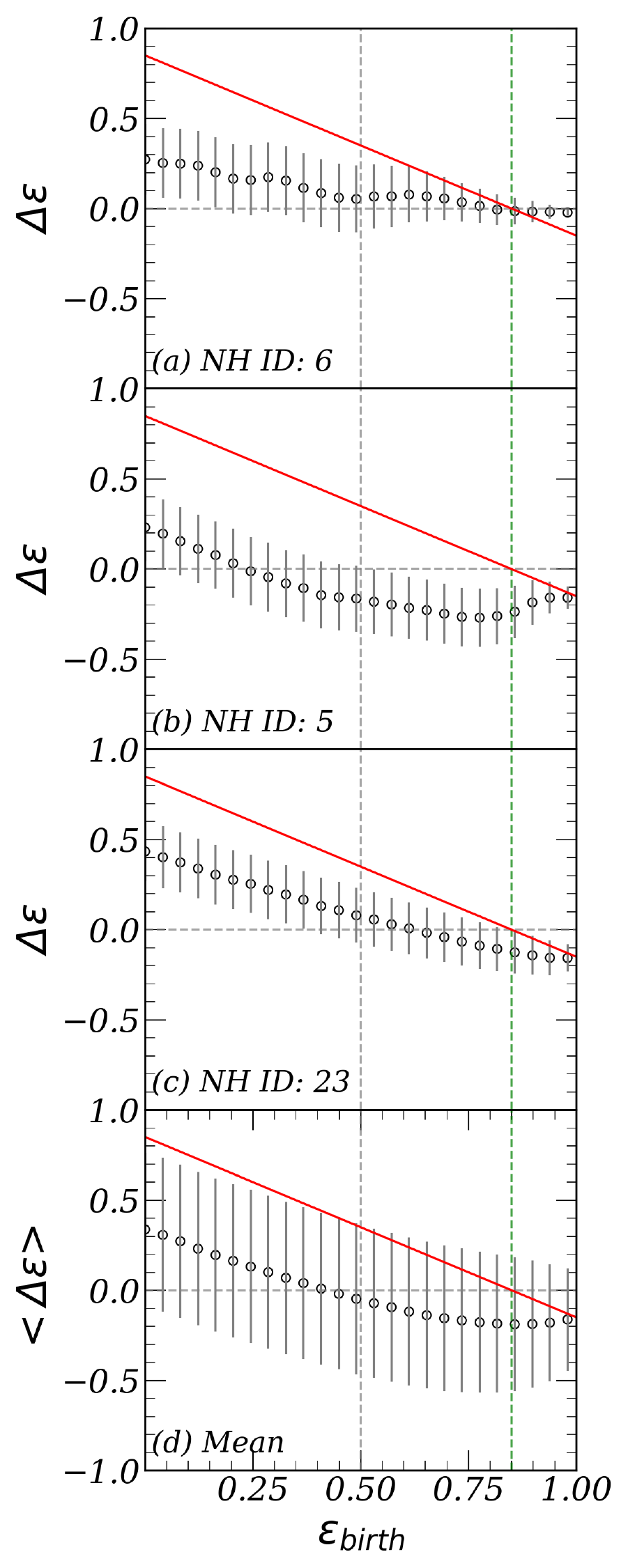}
\caption{
The mean change of circularity of the star particles between their birth and now ($\Delta \epsilon = \epsilon_{\rm now} - \epsilon_{\rm birth}$). Panels (a), (b), and (c) are for three sample galaxies, and Panel (d) shows the mean values for the whole sample of the 27 NH galaxies. 
The demarcation lines for the thin and thick disk components are marked by vertical dashed lines. 
The slanted line shows the criterion above which the star in the bin becomes kinematically thin.
Considering that most of the stars are born with a high value of circularity, most of the disk stars become kinematically hotter compared to their birth. 
} 
\label{fig_delcirc}
\end{figure}

Figure~\ref{fig_delcirc} shows the mean change of the circularity of the star particles binned by their birth circularity.
The top three panels show the values for three different galaxies, and the last panel shows the mean values for the whole sample of 27 NH galaxies.
The two vertical lines show the rough demarcation for the thin and thick disk components.
The slanted line shows the reference: a circularity change below this line will make the star particle kinematically thick ($\epsilon < 0.85$).
The first three panels highlight that the changes in the kinematic properties vary substantially from galaxy to galaxy, and it is important to have a large sample to obtain a general picture.
We then discuss the mean properties shown in Panel-(d).
It is useful to remember that most of the stars of the NH disk galaxies are born as kinematically thin, as illustrated in Figure~\ref{fig_circ_birthnow}.
The circularity changes of the stars born as kinematically-thin (four points of $\epsilon \geq 0.85$) are all negative, meaning that most of them become kinematically-thicker with time.
The magnitude of the change is so large that all points are below the red reference line, meaning that a large fraction of the thin-born stars becomes thick in the last snapshot.
Many of kinematically-hot stars (e.g., $\epsilon \lesssim 0.35$) become kinematically colder, probably due to the resonance with the galactic rotation. 
However, it should be noted that the mass fraction of stars born with such low circularity values is very small (see Figure~\ref{fig_circ_birthnow}).

The (ensemble average) stack in Panel-(d) is of clear dynamical interest as it corresponds to a (finite time) proxy  of a drift coefficient for circularity. 
Such a coefficient can be predicted from the first principle within the context of kinetic theory \citep{Binney1988}. 
Kinetic theory captures the fact that environmentally or internally driven force fluctuations operate on stars moving to zeroth order along their unperturbed orbit, and induce a slow resonant distortion of their trajectory: the mean orbital structure diffuses on secular timescales. 
Eventually one would need to compute all drift (vector) and (tensor) diffusion coefficients, which is beyond the scope of this paper.

\begin{figure}
\centering
\includegraphics[width=0.4\textwidth]{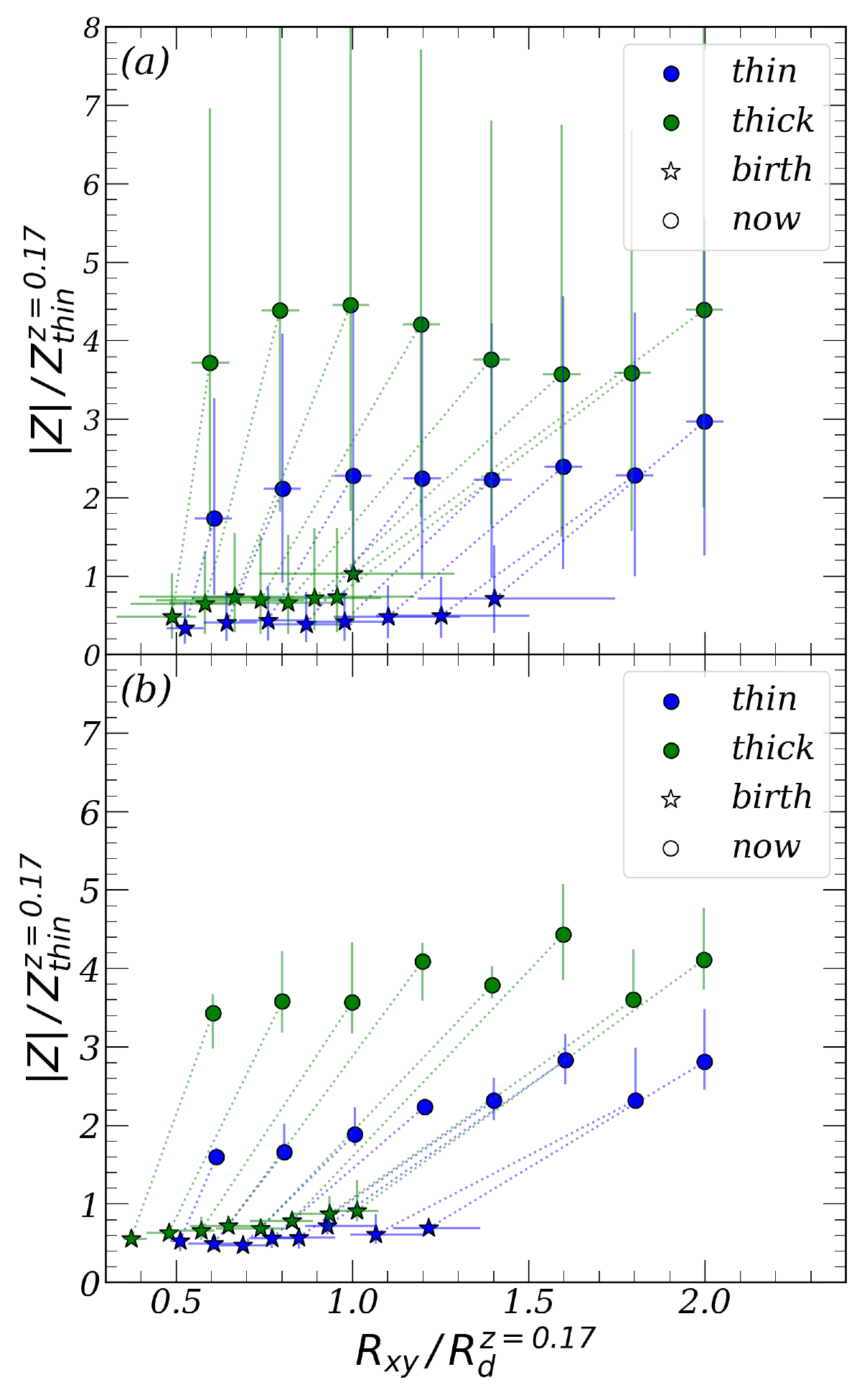}
\caption{
The same as Figure~\ref{fig_birthplace} but for the kinematically-thin ($\bar{\epsilon} > 0.85$) and -thick disk ($0.5 < \bar{\epsilon} \leq 0.85$) stars based on the $\epsilon =0.85$ cut. 
(a) The case of the galaxy with an \afe\ bimodality as shown in Figure~\ref{fig_chemical} and Figure~\ref{fig_morph_chem} (NH2 ID: 2).
(b) The mean of the 10 NH2 disk galaxies.
}
\label{fig_birthplace_kin}
\end{figure}

Figure~\ref{fig_birthplace_kin} shows the birthplaces of the kinematically-thin and -thick disk stars sampled by the $\epsilon =0.85$ cut (between thin and thick).
Panel-(a) shows the case of the NH2 galaxy with \afe\ bimodality, which was presented in Figures~\ref{fig_chemical} and \ref{fig_morph_chem}.
We recall that the Vintergatan galaxy suggested that its thick disk stars formed closer to the galactic center than its thin disk counterpart \citep{Agertz2021}.
In the case of NH2 galaxy, the kinematically-thick disk stars were born systematically closer to the galactic center and farther from the mid-plane than the kinematically-thin disk stars; but the difference is extremely small.
Although it is not shown here, when we inspect the stars that were born between 5 and 7 Gyr of lookback time during the minor merger as we mentioned in Section~\ref{chemical}, they were born near the galactic mid-plane, too.
We suspect that the differences in the details of the merger geometry and galaxy properties caused the differences between the Vitergatan and the NH2 simulations.
Panel-(b) shows the mean properties of 10 NH2 disk galaxies.
The birth positions of thin and thick disk stars are almost indistinguishable. 
Whether we use GMM (not shown here) or single-$\epsilon$ cut to select thin and thick disk stars does not make a notable difference in this diagram.

We summarise this section as follows. 
We detected kinematic groups/components based on the kinematic properties of the star particles in the simulation. 
The use of reasonable criteria for energy and angular momentum successfully detected thin and thick disk particles; however, the separation remained somewhat arbitrary. 
If we use $\epsilon =0.85$ as one of the critical conditions, the mean value of the thick disk fraction in $r$-band luminosity is $34\pm 15$\% or $48 \pm 13$\% in mass, which is consistent with the observations.
Orbital properties are not well traced by the alpha abundance.
Kinematic information, such as energy and angular momentum, is useful for studying the physical processes that generated the orbital properties of stars using theoretical/simulation models. 
However, it is not trivial to measure them observationally.
Fortunately, the vertical luminosity profile fits provide key kinematic properties of thick disks reasonably closely.
Most stars are born near the galactic plane and get dynamically heated with time.
Thick disks are naturally generated as an ensemble of stars that are heated for different lengths of time after their birth.

\section{Conclusions}
\label{conclusion}

We give a brief summary first. 
We used the NewHorizon simulation in this investigation. 
With its high resolutions in space and mass, we resolved the thin disks of massive spiral galaxies, which makes our investigation of thick disks possible.
NH also has a high temporal resolution which allows accurate measurements of the kinematic properties of star particles. 
In addition, we used the NewHorizon2 simulation which further provides the evolution of nine chemical elements.
This allows a comparison between the results from different approaches that are used to estimate the significance of thick disks.
To ensure that the models reasonably resolve the thin disks, we used only the massive ($M_* \geq 10^{10}$~\msun).
In addition, we used a stellar $v/\sigma > 1$ cut to select disk galaxies.
We have 27 galaxies from NH and 10 galaxies from NH2 in our sample. 

We first performed vertical profile fitting in $r$-band using up to three $\mathrm{sech}^2$ components with different scale heights.
Based on the BIC statistical assessment, 0, 22, and 5 galaxies were found to prefer one, two, or three components, respectively.
This may have a profound implication for the study of thick disks. 
The majority of the stars in spiral galaxies are probably born within a thin cold disk and get gradually heated with time through many processes.
If there were $n$ rather distinctly more active periods of star formation in the galactic disk either via secular regulation, cyclic gas inflow, or episodic merger events, we may find $n$ disk components with different scale heights, not necessarily because the $n$ components were born with a different scale height but more likely because they were kinematically heated for different lengths of time.
Therefore, it is not surprising to find more than two disk components in some galaxies.
This has indeed been addressed in the observational study of \citet{Comeron2018}.
The vertical scale heights of the thin and thick disks and the luminosity contribution of the thick disks of the NH model galaxies were found to be in reasonable agreement with the observed data of external edge-on galaxies.

Assuming that the vertical profile fit gives  reliable information regarding the distribution of thin and thick disk stars, we first attempted to sample thick disk stars based on position. 
We sampled spatially-thin disk stars near the galactic mid-plane and spatially-thick disk stars away from the mid-plane.
A strong advantage of this sampling technique is that it could be compared with the observed data of external galaxies easily. 
A disadvantage on the other hand is that both spatially-thin and -thick disk stars sampled this way are bound to be contaminated by each other.
We checked where these stars were born initially.
We confirm the earlier result of \citet{Park2021} that both thin and thick disk stars were born near the mid-plane.
Their birthplaces were almost indistinguishable.

We then sampled thin and thick disk stars based on alpha abundance, \afe.
Some observational studies have suggested a connection between the kinematic properties and \afe\ properties of disk stars in the case of the Milky Way.
The bimodality in \afe\ is often considered as evidence for the distinct origin of the thick disk from the thin disk.
Besides, a naive projection of the \afe\ bimodality would indicate a substantially higher prominence of the thick disk than that suggested by profile fit studies.
The NH2 galaxies showed a variety in the pattern of \afe\ distribution, while there usually was a good correlation between age and \afe.
The NH simulation (and thus NH2 as well) is for an average field environment.
The majority of its galaxies have not undergone major mergers since they developed a galactic scale disk or since $z \approx 2$.
In such an effectively ``closed box'' of chemical evolution testbed, \afe\ by and large monotonically decreases with time.
Thus, the \afe\ bimodality, more specifically the high \afe\ peak hints for a separate event of star formation from the main star formation history of the Milky Way disk. 
We found a galaxy in our sample that resembles the Milky Way in terms of \afe\ bimodality.
The high \afe\ peak stars were indeed born during the high star formation phase caused by a gas-rich galaxy merger accretion 6 Gyr ago. 
The high \afe\ peak found in the Milky Way disk may have had a similar origin. 

In the NH2 galaxy with an \afe\ bimodality, the stars in the low and high-\afe\ peaks roughly appeared to be thin and thick in spatial distribution at first glance.
But this correspondence was found to be weak by the following kinematic analysis.
The low-\afe-peak stars were indeed dominantly kinematically-thin disk stars following the coherent rotation of the disk.
However, the high-\afe-peak stars were not just kinematically-thick disk stars but were significantly contaminated by the stars of other kinematic properties (such as thin disk stars or dispersion-dominant stars).

We used the unsupervised machine learning technique, Gaussian Mixture Model, to detect kinematical groups of stars in each galaxy using energy and angular momentum. 
It was reasonably straightforward to identify the most significant components in the simulated galaxy.
Whether GMM-detected components are  ``distinct'' is debatable; yet, it is probably more reliable than applying a single value cut of any kind.
When we classified the particles using reasonable conditions of kinematic properties, we found good correspondence between the kinematic decomposition and profile-fit decomposition of thin and thick disks.
Kinematically-selected thin-disk stars were usually low in \afe\ and young in age, hence good correspondence. 
However, the thick-disk counterpart did not represent any single dominant group of stars in age or chemical properties.
This is because, while today's kinematically-thin disk stars are predominantly young stars bearing similar chemical properties,  the thick disk counterpart is a cumulative population of stars that were born at different epochs with different chemical backgrounds.

Applying an arbitrary yet reasonable criterion to the kinematic data of the 27 NH galaxies indicated that $34\pm 15$\% in $r$-band luminosity or $48 \pm 13$\% in mass of the disk was found to be kinematically hot.
These measurements are in reasonable agreement with the observational data of edge-on disk galaxies and the Milky Way.

The presence of thick disks is probably not surprising after all.
Star formation in a disk galaxy that is not involved in a significant galaxy interaction probably occurs predominantly within the thin cold gas disk.
So the youngest stars likely share similar kinematic properties to the galactic gas disk, rotating fast.
The stars are kinematically heated by force fluctuations along their orbit.
Perturbation causes an increase of entropy and thus must build up with events and with time.
If the sources of perturbation were numerous and effectively smooth, disk thickening would likely be smooth.
If the dominant source of perturbation was $n$ dramatic events such as galaxy mergers, their signature will be $n$ discrete thick disk populations. 
If only one important merger happened in the detectable past, it may show just one thick disk component, which could be the case of the Milky Way disk.
However, this does not necessarily mean that the merger was the dominant contributor to the Milky Way thick disk. 
It is probably more likely that its added contribution made the thick disk more conspicuous.
Depending on the detailed nature of the perturbation events, a galaxy may have one dominant thin disk, two disks, or $n$ disk components.

Eventually one should aim to ensemble average orbital diffusion in 3+1 D orbital and chemical space, in the spirit of the bottom panel of Figure~\ref{fig_delcirc}, so as to predict the mean effect of such mixture of cosmic accretion event. 
This could then be compared to inhomogenous kinetic theory \citep{Fouvry2017} which predicts such diffusion rate.
The GMM would prove extremely useful in this context.

\section*{Acknowledgements}

Y.D., S.K.Y., and S.H. ran the NewHorizon and NewHorizon2 simulations.
J.R. and T.K. implemented the chemical evolution in the NewHorizon2 simulation. 
JK.J. performed most of the hands-on analysis, and S.K.Y. wrote the draft.
C.P., J.D., S.P., K.K., and M.P. contributed to the completion of the manuscript. 
We thank Young Sun Lee for making the SDSS data available to us and for useful discussion.
We also thank Seyoung Jeon for his careful proofreading.

This work was granted access to the HPC resources of CINES under the allocations  c2016047637, A0020407637 and A0070402192 by Genci, KSC-2017-G2-0003, KSC-2020-CRE-0055 and KSC-2020-CRE-0280 by KISTI, and as a “Grand Challenge” project granted by GENCI on the AMD Rome extension of the Joliot Curie supercomputer at TGCC.
The large data transfer was supported by KREONET which is managed and operated by KISTI. 
S.K.Y. acknowledges support from the Korean National Research Foundation (NRF-2020R1A2C3003769). 
T.K. was supported in part by the National Research Foundation of Korea (NRF-2020R1C1C1007079).
J.R. was supported by the KASI-Yonsei Postdoctoral Fellowship and was supported by the Korea Astronomy and Space Science Institute under the R\&D program (Project No. 2023-1-830-00), supervised by the Ministry of Science and ICT.
This study was funded in part by the NRF-2022R1A6A1A03053472 grant and the BK21Plus program.
Part of the drafting of the paper was carried out during S.K.Y's participation in the KITP workshop on Cosmic Web (Feb. 2023) which was supported in part by the National Science Foundation under Grant No. NSF PHY-1748958.
\\
\vspace{-0.5cm}

\bibliography{references}{}

\begin{thebibliography}{}
\expandafter\ifx\csname natexlab\endcsname\relax\def\natexlab#1{#1}\fi
\providecommand{\url}[1]{\href{#1}{#1}}

\bibitem[{Abadi {et~al.}(2003)Abadi, Navarro, Steinmetz, \&
  Eke}]{Abadi2003SimulationsDisks}
Abadi, M.~G., Navarro, J.~F., Steinmetz, M., \& Eke, V.~R. 2003, \apj, 597, 21

\bibitem[{{Agertz} {et~al.}(2021){Agertz}, {Renaud}, {Feltzing}, {Read},
  {Ryde}, {Andersson}, {Rey}, {Bensby}, \& {Feuillet}}]{Agertz2021}
{Agertz}, O., {Renaud}, F., {Feltzing}, S., {et~al.} 2021, \mnras, 503, 5826

\bibitem[{Asplund {et~al.}(2009)Asplund, Grevesse, Sauval, \&
  Scott}]{Asplund2009TheSun}
Asplund, M., Grevesse, N., Sauval, A.~J., \& Scott, P. 2009, \araa, 47, 481

\bibitem[{{Bensby} \& {Feltzing}(2006)}]{Bensby2006}
{Bensby}, T., \& {Feltzing}, S. 2006, \mnras, 367, 1181

\bibitem[{{Bensby} {et~al.}(2014){Bensby}, {Feltzing}, \& {Oey}}]{Bensby2014}
{Bensby}, T., {Feltzing}, S., \& {Oey}, M.~S. 2014, \aap, 562, A71

\bibitem[{{Binney} \& {Lacey}(1988)}]{Binney1988}
{Binney}, J., \& {Lacey}, C. 1988, \mnras, 230, 597

\bibitem[{Bland-Hawthorn \& Gerhard(2016)}]{Bland-Hawthorn2016TheProperties}
Bland-Hawthorn, J., \& Gerhard, O. 2016, \araa, 54, 529

\bibitem[{Bovy \& Rix(2013)}]{Bovy2013AKpc}
Bovy, J., \& Rix, H.~W. 2013, \apj, 779, 115

\bibitem[{Brook {et~al.}(2004)Brook, Kawata, Gibson, \&
  Freeman}]{Brook2004TheUniverse}
Brook, C.~B., Kawata, D., Gibson, B.~K., \& Freeman, K.~C. 2004, \apj, 612, 894

\bibitem[{{Buder} {et~al.}(2021){Buder}, {Sharma}, {Kos}, {Amarsi},
  {Nordlander}, {Lind}, {Martell}, {Asplund}, {Bland-Hawthorn}, {Casey}, {de
  Silva}, {D'Orazi}, {Freeman}, {Hayden}, {Lewis}, {Lin}, {Schlesinger},
  {Simpson}, {Stello}, {Zucker}, {Zwitter}, {Beeson}, {Buck}, {Casagrande},
  {Clark}, {{\v{C}}otar}, {da Costa}, {de Grijs}, {Feuillet}, {Horner},
  {Kafle}, {Khanna}, {Kobayashi}, {Liu}, {Montet}, {Nandakumar}, {Nataf},
  {Ness}, {Spina}, {Tepper-Garc{\'\i}a}, {Ting}, {Traven},
  {Vogrin{\v{c}}i{\v{c}}}, {Wittenmyer}, {Wyse}, {{\v{Z}}erjal}, \& {Galah
  Collaboration}}]{Buder2021_GALAH_DR3}
{Buder}, S., {Sharma}, S., {Kos}, J., {et~al.} 2021, \mnras, 506, 150

\bibitem[{{Burstein}(1979)}]{Burstein1979}
{Burstein}, D. 1979, \apj, 234, 829

\bibitem[{{Chabrier}(2005)}]{Chabrier05}
{Chabrier}, G. 2005, in Astrophysics and Space Science Library, Vol. 327, The
  Initial Mass Function 50 Years Later, ed. E.~{Corbelli}, F.~{Palla}, \&
  H.~{Zinnecker}, 41

\bibitem[{{Comer{\'o}n} {et~al.}(2018){Comer{\'o}n}, {Salo}, \&
  {Knapen}}]{Comeron2018}
{Comer{\'o}n}, S., {Salo}, H., \& {Knapen}, J.~H. 2018, \aap, 610, A5

\bibitem[{Comer{\'{o}}n {et~al.}(2011)Comer{\'{o}}n, Elmegreen, Knapen, Salo,
  Laurikainen, Laine, Athanassoula, Bosma, Sheth, Regan, Hinz, De~Paz,
  Men{\'{e}}ndez-Delmestre, Mizusawa, Mũoz-Mateos, Seibert, Kim, Elmegreen,
  Gadotti, Ho, Holwerda, Lappalainen, Schinnerer, \&
  Skibba}]{Comeron2011ThickBaryons}
Comer{\'{o}}n, S., Elmegreen, B.~G., Knapen, J.~H., {et~al.} 2011, \apj, 741,
  28

\bibitem[{{Du} {et~al.}(2020){Du}, {Ho}, {Debattista}, {Pillepich}, {Nelson},
  {Zhao}, \& {Hernquist}}]{Du2020}
{Du}, M., {Ho}, L.~C., {Debattista}, V.~P., {et~al.} 2020, \apj, 895, 139

\bibitem[{{Du} {et~al.}(2019){Du}, {Ho}, {Zhao}, {Shi}, {Debattista},
  {Hernquist}, \& {Nelson}}]{Du2019}
{Du}, M., {Ho}, L.~C., {Zhao}, D., {et~al.} 2019, \apj, 884, 129

\bibitem[{{Dubois} {et~al.}(2021){Dubois}, {Beckmann}, {Bournaud}, {Choi},
  {Devriendt}, {Jackson}, {Kaviraj}, {Kimm}, {Kraljic}, {Laigle}, {Martin},
  {Park}, {Peirani}, {Pichon}, {Volonteri}, \& {Yi}}]{Dubois2021}
{Dubois}, Y., {Beckmann}, R., {Bournaud}, F., {et~al.} 2021, \aap, 651, A109

\bibitem[{{Fouvry} {et~al.}(2017){Fouvry}, {Pichon}, {Chavanis}, \&
  {Monk}}]{Fouvry2017}
{Fouvry}, J.-B., {Pichon}, C., {Chavanis}, P.-H., \& {Monk}, L. 2017, \mnras,
  471, 2642

\bibitem[{{Gilmore} \& {Reid}(1983)}]{Gilmore1983}
{Gilmore}, G., \& {Reid}, N. 1983, \mnras, 202, 1025

\bibitem[{{Halle} {et~al.}(2018){Halle}, {Di Matteo}, {Haywood}, \&
  {Combes}}]{Halle2018}
{Halle}, A., {Di Matteo}, P., {Haywood}, M., \& {Combes}, F. 2018, \aap, 616,
  A86

\bibitem[{{Iwamoto} {et~al.}(1999){Iwamoto}, {Brachwitz}, {Nomoto},
  {Kishimoto}, {Umeda}, {Hix}, \& {Thielemann}}]{Iwamoto99}
{Iwamoto}, K., {Brachwitz}, F., {Nomoto}, K., {et~al.} 1999, \apjs, 125, 439

\bibitem[{{Jang} {et~al.}(2022){Jang}, {Yi}, {Dubois}, {Rhee}, {Pichon},
  {Kimm}, {Devriendt}, {Volonteri}, {Kaviraj}, {Peirani}, {Oh}, \&
  {Croom}}]{Jang2023}
{Jang}, J.~K., {Yi}, S.~K., {Dubois}, Y., {et~al.} 2022, arXiv e-prints,
  arXiv:2211.00931

\bibitem[{{J{\"o}nsson} {et~al.}(2020){J{\"o}nsson}, {Holtzman}, {Allende
  Prieto}, {Cunha}, {Garc{\'\i}a-Hern{\'a}ndez}, {Hasselquist}, {Masseron},
  {Osorio}, {Shetrone}, {Smith}, {Stringfellow}, {Bizyaev}, {Edvardsson},
  {Majewski}, {M{\'e}sz{\'a}ros}, {Souto}, {Zamora}, {Beaton}, {Bovy}, {Donor},
  {Pinsonneault}, {Poovelil}, \& {Sobeck}}]{Jonsson2020_APOGEE_DR16}
{J{\"o}nsson}, H., {Holtzman}, J.~A., {Allende Prieto}, C., {et~al.} 2020, \aj,
  160, 120

\bibitem[{Juric {et~al.}(2008)Juric, Ivezic, Brooks, Lupton, Schlegel,
  Finkbeiner, Padmanabhan, Bond, Sesar, Rockosi, Knapp, Gunn, Sumi, Schneider,
  Barentine, Brewington, Brinkmann, Fukugita, Harvanek, Kleinman, Krzesinski,
  Long, Neilsen, Nitta, Snedden, \& York}]{Juric2008TheDistribution}
Juric, M., Ivezic, Z., Brooks, A., {et~al.} 2008, \apj, 673, 864

\bibitem[{Kazantzidis {et~al.}(2008)Kazantzidis, Bullock, R., Kravtsov, \&
  Moustakas}]{Kazantzidis2008ColdAccretion}
Kazantzidis, S., Bullock, J.~S., R., Z.~A., Kravtsov, A.~V., \& Moustakas,
  L.~A. 2008, \apj, 688, 254

\bibitem[{Kimm {et~al.}(2017)Kimm, Katz, Haehnelt, Rosdahl, Devriendt, \&
  Slyz}]{Kimm2017Feedback-regulatedReionisation}
Kimm, T., Katz, H., Haehnelt, M., {et~al.} 2017, \mnras, 466, 4826

\bibitem[{{Kobayashi} {et~al.}(2006){Kobayashi}, {Umeda}, {Nomoto}, {Tominaga},
  \& {Ohkubo}}]{Kobayashi06}
{Kobayashi}, C., {Umeda}, H., {Nomoto}, K., {Tominaga}, N., \& {Ohkubo}, T.
  2006, \apj, 653, 1145

\bibitem[{Lee {et~al.}(2011)Lee, Beers, An, Ivezi{\'{c}}, Just, Rockosi,
  Morrison, Johnson, Sch{\"{o}}nrich, Bird, Yanny, Harding, \&
  Rocha-Pinto}]{Lee2011FormationSample}
Lee, Y.~S., Beers, T.~C., An, D., {et~al.} 2011, \apj, 738, 187

\bibitem[{{Leitherer} {et~al.}(2014){Leitherer}, {Ekstr{\"o}m}, {Meynet},
  {Schaerer}, {Agienko}, \& {Levesque}}]{Leitherer14}
{Leitherer}, C., {Ekstr{\"o}m}, S., {Meynet}, G., {et~al.} 2014, \apjs, 212, 14

\bibitem[{{Leitherer} {et~al.}(1999){Leitherer}, {Schaerer}, {Goldader},
  {Delgado}, {Robert}, {Kune}, {de Mello}, {Devost}, \&
  {Heckman}}]{Leitherer1999}
{Leitherer}, C., {Schaerer}, D., {Goldader}, J.~D., {et~al.} 1999, \apjs, 123,
  3

\bibitem[{{Maeder} \& {Meynet}(2000)}]{Maeder00}
{Maeder}, A., \& {Meynet}, G. 2000, \aap, 361, 159

\bibitem[{{Maoz} {et~al.}(2012){Maoz}, {Mannucci}, \& {Brandt}}]{Maoz12}
{Maoz}, D., {Mannucci}, F., \& {Brandt}, T.~D. 2012, \mnras, 426, 3282

\bibitem[{{Mart{\'\i}nez-Lombilla} \& {Knapen}(2019)}]{Martinez-Lombilla2019}
{Mart{\'\i}nez-Lombilla}, C., \& {Knapen}, J.~H. 2019, \aap, 629, A12

\bibitem[{{Norris} \& {Ryan}(1991)}]{Norris1991}
{Norris}, J.~E., \& {Ryan}, S.~G. 1991, \apj, 380, 403

\bibitem[{{Park} {et~al.}(2021){Park}, {Yi}, {Peirani}, {Pichon}, {Dubois},
  {Choi}, {Devriendt}, {Kaviraj}, {Kimm}, {Kraljic}, \& {Volonteri}}]{Park2021}
{Park}, M.~J., {Yi}, S.~K., {Peirani}, S., {et~al.} 2021, \apjs, 254, 2

\bibitem[{Peebles(1969)}]{Peebles1969OriginGalaxies}
Peebles, P. J.~E. 1969, \apj, 155, 393

\bibitem[{Pichon {et~al.}(2011)Pichon, Pogosyan, Kimm, Slyz, Devriendt, \&
  Dubois}]{Pichon2011}
Pichon, C., Pogosyan, D., Kimm, T., {et~al.} 2011, \mnras, 418, 2493

\bibitem[{Quinn {et~al.}(1993)Quinn, Hernquist, \&
  Fullagar}]{Quinn1993HeatingMergers}
Quinn, P.~J., Hernquist, L., \& Fullagar, D.~P. 1993, \apj, 403, 74

\bibitem[{{Rasera} \& {Teyssier}(2006)}]{Rasera2006}
{Rasera}, Y., \& {Teyssier}, R. 2006, \aap, 445, 1

\bibitem[{{Reddy} {et~al.}(2006){Reddy}, {Lambert}, \& {Allende
  Prieto}}]{Reddy2006}
{Reddy}, B.~E., {Lambert}, D.~L., \& {Allende Prieto}, C. 2006, \mnras, 367,
  1329

\bibitem[{Ro{\v{s}}kar {et~al.}(2013)Ro{\v{s}}kar, Debattista, \&
  Loebman}]{Roskar2013TheDiscs}
Ro{\v{s}}kar, R., Debattista, V.~P., \& Loebman, S.~R. 2013, \mnras, 433, 976

\bibitem[{{Schaller} {et~al.}(1992){Schaller}, {Schaerer}, {Meynet}, \&
  {Maeder}}]{Schaller92}
{Schaller}, G., {Schaerer}, D., {Meynet}, G., \& {Maeder}, A. 1992, \aaps, 96,
  269

\bibitem[{{Sellwood} \& {Carlberg}(1984)}]{SellwoodJ.A.1984SpiralFormation}
{Sellwood}, J.~A., \& {Carlberg}, R.~G. 1984, \apj, 282, 61

\bibitem[{{Sellwood} {et~al.}(1998){Sellwood}, {Nelson}, \&
  {Tremaine}}]{sellwood1998}
{Sellwood}, J.~A., {Nelson}, R.~W., \& {Tremaine}, S. 1998, \apj, 506, 590

\bibitem[{S{\'{e}}rsic(1963)}]{Sersic1963InfluenceGalaxy}
S{\'{e}}rsic, J.~L. 1963, BAAA, 6, 41

\bibitem[{Spitzer \& Schwarzschild(1951)}]{Spitzer1951TheVelocities}
Spitzer, L., \& Schwarzschild, M. 1951, 114, 385

\bibitem[{Teyssier(2002)}]{Teyssier2002CosmologicalRefinement}
Teyssier, R. 2002, \aap, 385, 337

\bibitem[{{Tinsley}(1980)}]{Tinsley1980}
{Tinsley}, B.~M. 1980, \fcp, 5, 287

\bibitem[{{van der Kruit}(1981)}]{vanderKruit1981}
{van der Kruit}, P.~C. 1981, \aap, 99, 298

\bibitem[{{Vera-Ciro} {et~al.}(2014){Vera-Ciro}, {D'Onghia}, {Navarro}, \&
  {Abadi}}]{Vera-Ciro2014heating}
{Vera-Ciro}, C., {D'Onghia}, E., {Navarro}, J., \& {Abadi}, M. 2014, \apj, 794,
  173

\bibitem[{{Woosley} \& {Weaver}(1995)}]{Woosley95}
{Woosley}, S.~E., \& {Weaver}, T.~A. 1995, \apjs, 101, 181

\bibitem[{{Yoachim} \& {Dalcanton}(2006)}]{Yoachim2006StructuralGalaxies}
{Yoachim}, P., \& {Dalcanton}, J.~J. 2006, \aj, 131, 226

\bibitem[{Yoachim \& Dalcanton(2008)}]{Yoachim2008TheGalaxies}
Yoachim, P., \& Dalcanton, J.~J. 2008, \apj, 682, 1004

\end{thebibliography}
\bibliographystyle{aasjournal}

\appendix
\restartappendixnumbering

\section{Appendix}
\label{sec appendix_A}
\begin{figure*}
\includegraphics[width=0.95\textwidth]{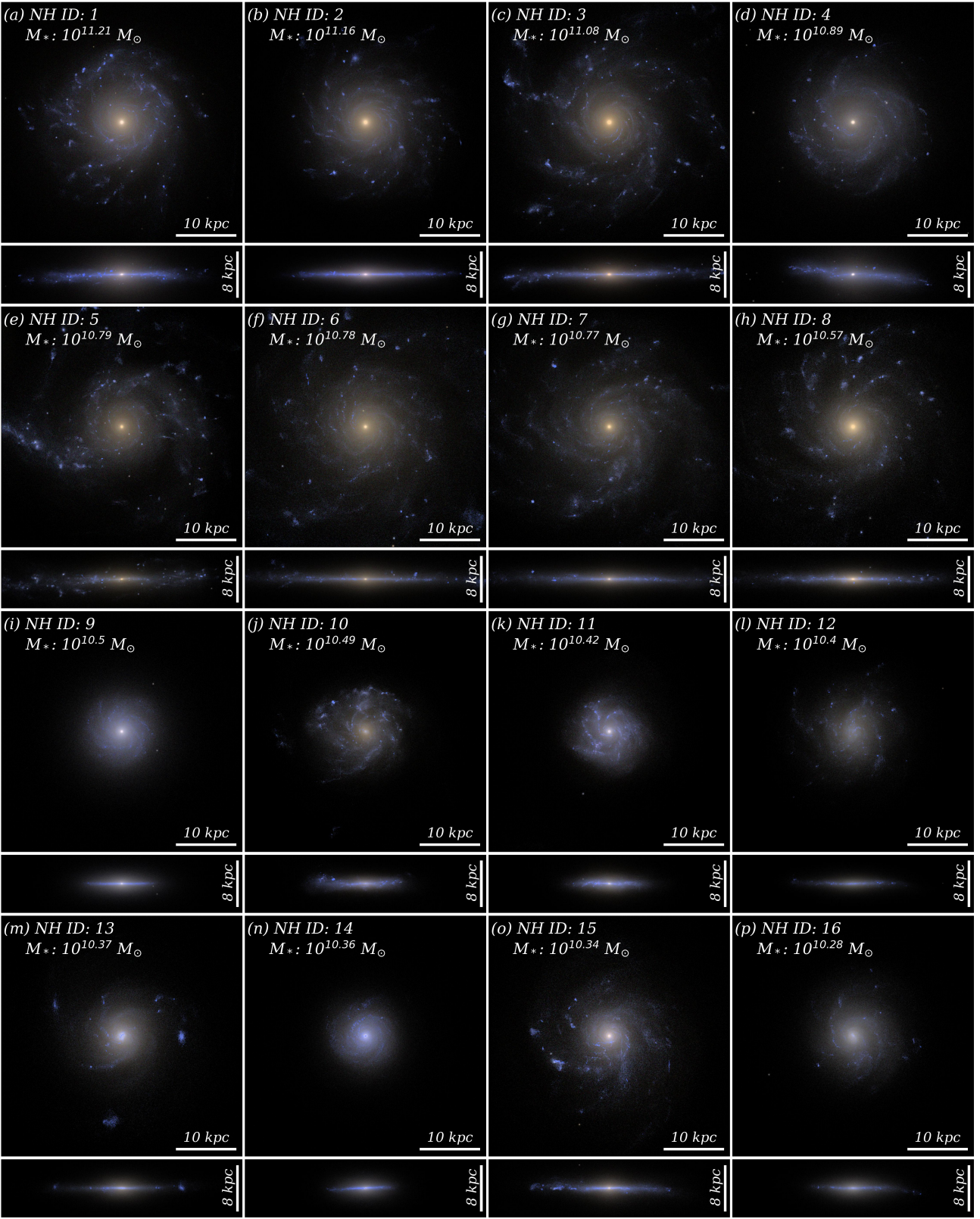}
\caption{
The face-on and edge-on images of the sixteen most massive disk galaxies in the NH simulation.
The blue, green, and red components in the mock images correspond to the Sloan Digital Sky Survey (SDSS) $g$, $r$, and $i$ bands.
}
\label{fig_appendixA1}
\end{figure*}

\begin{figure*}
\includegraphics[width=0.95\textwidth]{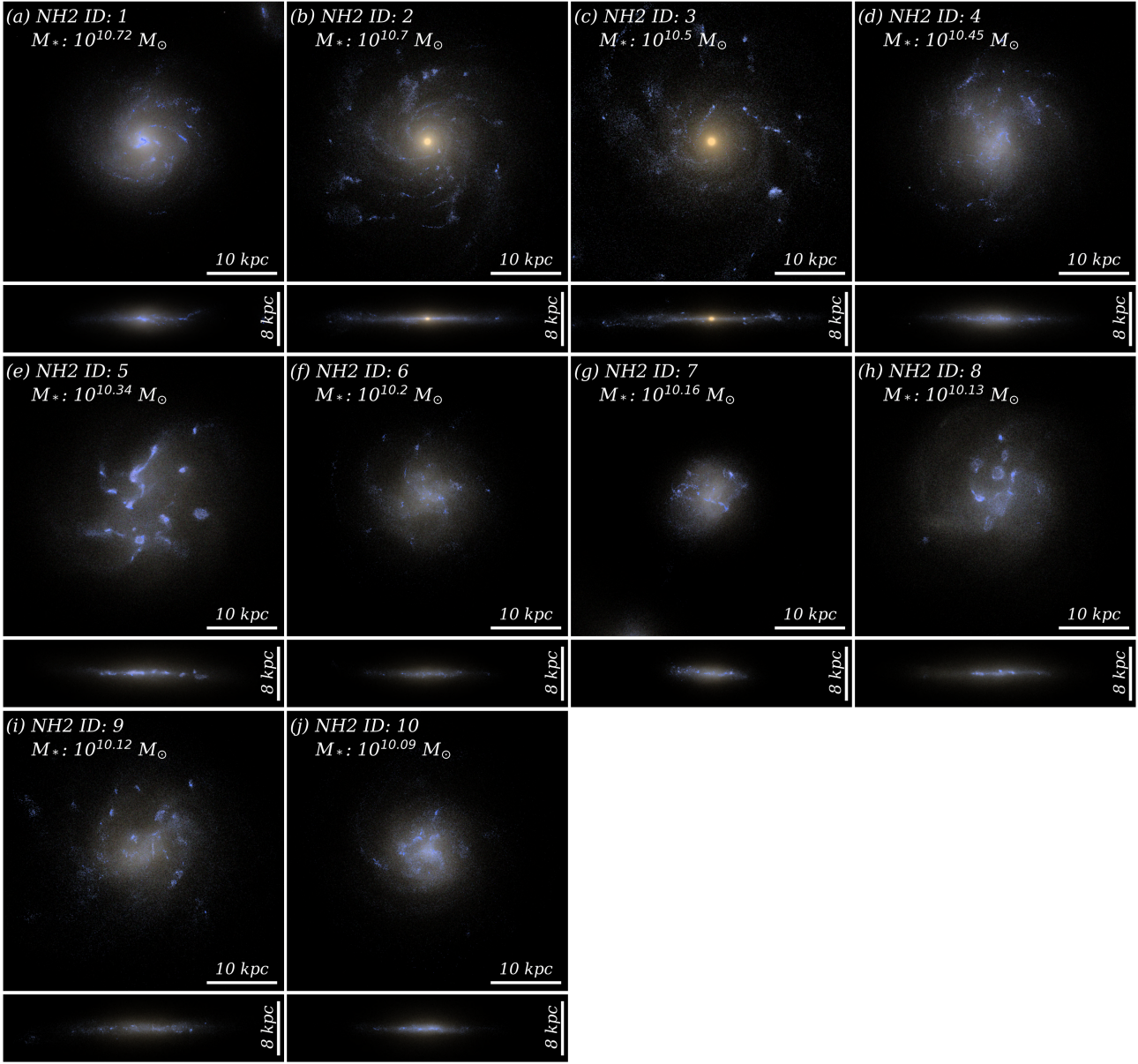}
\caption{
Same as Figure~\ref{fig_appendixA1} but for the sample of ten NH2 galaxies. 
}
\label{fig_appendixA2}
\end{figure*}

\begin{figure*}
\includegraphics[width=0.95\textwidth]{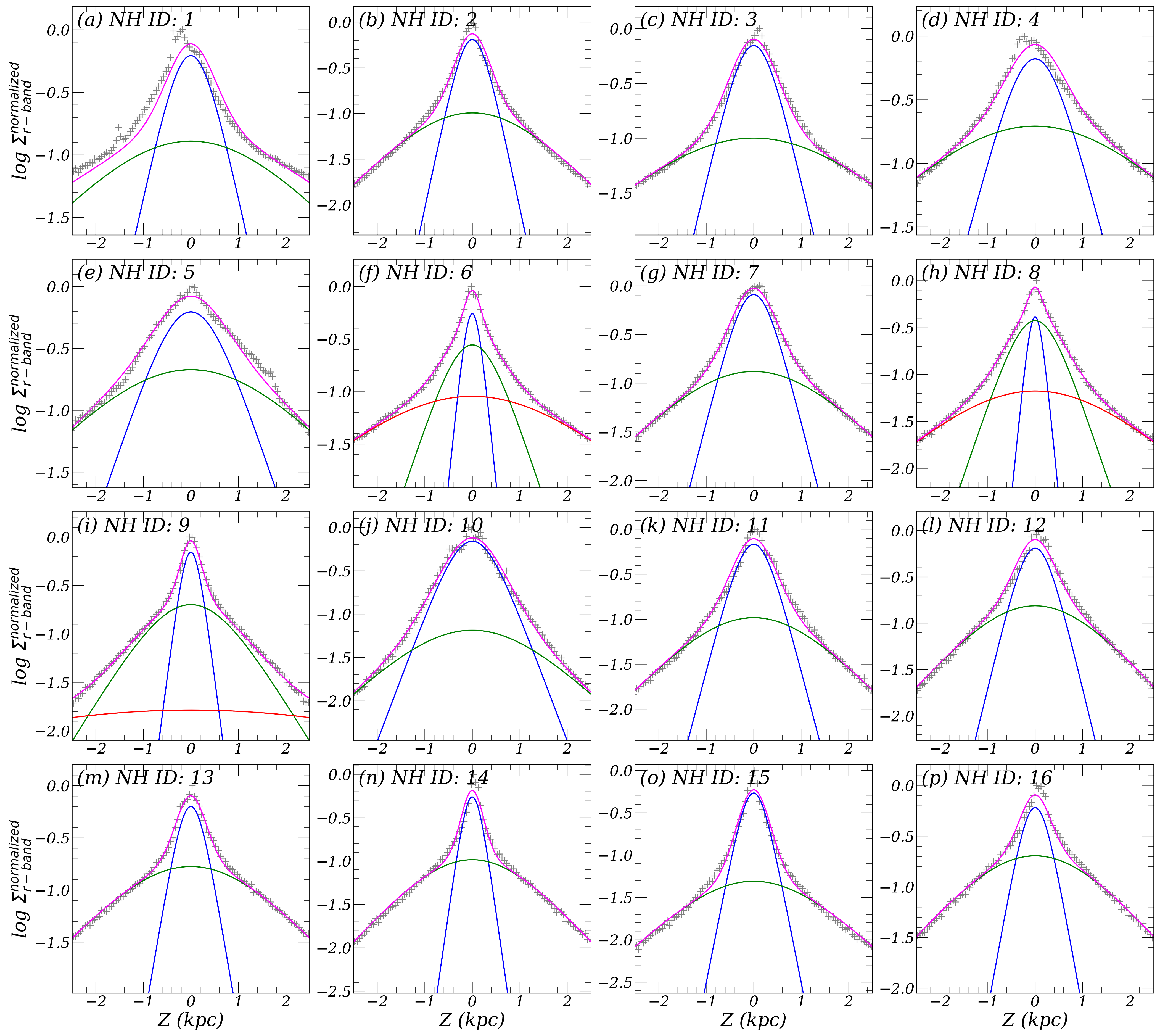}
\caption{
The BIC-selected vertical fitting result, measured at 1 $\pm$ 0.1$\,R_{d}$ for the same NH galaxies in Figure~\ref{fig_appendixA1}. 
The data points (gray plus), thin disk (blue line), thick disk (green line), extra component (red line), and total fit (magenta line) are presented.
} 
\label{fig_appendixA3}
\end{figure*}

\begin{figure*}
\includegraphics[width=0.95\textwidth]{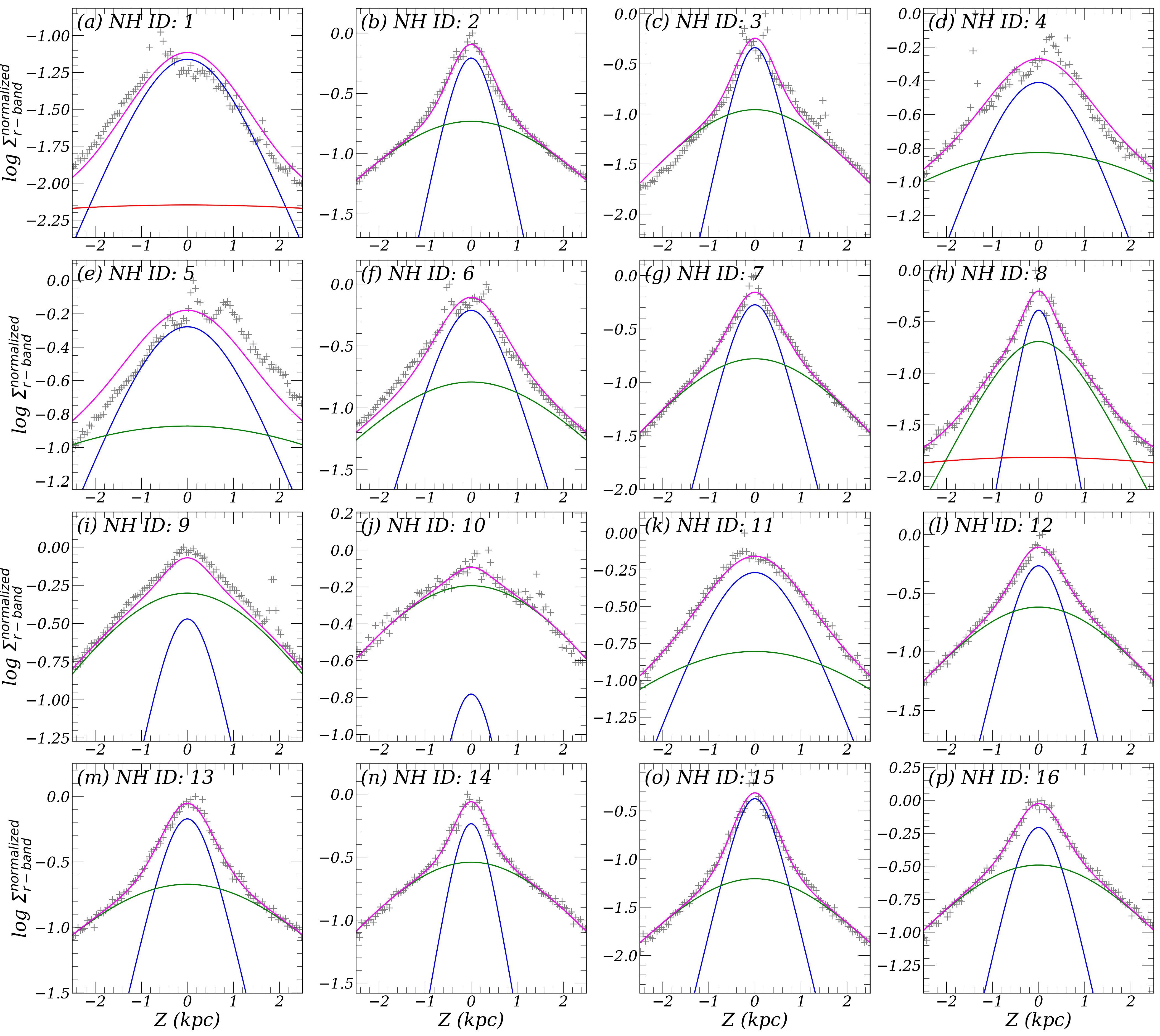}
\caption{
Same as Figure~\ref{fig_appendixA3} but for the measurement performed at 2 $\pm$ 0.1$\,R_{d}$ for reference. 
} 
\label{fig_appendixA4}
\end{figure*}

We provide supplementary information on the galaxies from the NewHorizon (NH) and NewHorizon2 (NH2) simulations in this section. They are not critical to the discussion presented in the main text but useful as a reference. 
Table~\ref{tab:Table1} provides the basic properties of the NH galaxies used or presented in this study.
Figures~\ref{fig_appendixA1} and \ref{fig_appendixA2} show the face-on and edge-on images of the sampled NH and NH2 galaxies, respectively. 
Figure~\ref{fig_appendixA3} shows the vertical profile fits to the NH galaxies in measured at the radial distance $1 \pm 0.1 R_{\rm d}$ from the galactic center, where $R_{\rm d}$ stands for the scale length of the disk, and the vertical distance $|z| \leq 3$~kpc from the galactic mid-plane.
Figure~\ref{fig_appendixA4} shows the same but for the measurement at $2 \pm 0.1 R_{\rm d}$ for comparison.
Figure~\ref{fig_appendixA5} shows the \afe\ vs. \feh\ maps of nine massive NH2 galaxies that appeared in Figure~\ref{fig_appendixA2}.
Figure~\ref{fig_appendixA6} shows the corresponding \afe\ vs. age maps to the galaxies presented in Figure~\ref{fig_appendixA5}.
Figures~\ref{fig_appendixA7} and \ref{fig_appendixA8} show the Gaussian Mixture Model (GMM) phase space diagrams in the same format as Figure~\ref{fig_gmm} in the main text but for additional NH galaxies. 
Figure~\ref{fig_appendixA9} shows \afe\ vs. circularity at present for the NH2 galaxies except those already shown in Figure~\ref{fig_alpha_circ} in the main text.
Figure~\ref{fig_appendixA10} shows the circularity at birth and at present for the sixteen NH galaxies shown in Figure~\ref{fig_appendixA1}.

\begin{table}[]
\centering
\caption{General properties of the NewHorizon galaxies}
\label{tab:Table1}
\begin{tabular*}{\columnwidth}{@{\extracolsep{\fill}} ccccccccccccc}
\hline 
\hline 
ID$^{a}$ & $M_{*,3R_{\rm e}}$$^{b}$ & $V_{\rm circ}$$^{c}$ & $R_{\rm e}$$^{d}$ & $R_{\rm d}$$^{e}$ & $R_{\rm 90}$$^{f}$ & \multicolumn{2}{c}{$z_{\rm thin}$$^{g}$} & \multicolumn{2}{c}{$z_{\rm thick}$$^{h}$} & \multicolumn{2}{c}{$(L_{\rm thick}/L_{\rm tot})$$^{i}$}& $(M_{\rm thick}/M_{\rm tot})$$^{j}$  \\ [5pt]
\hline
\multicolumn{1}{c}{} & \multicolumn{1}{c}{} & \multicolumn{1}{c}{} & \multicolumn{1}{c}{} & \multicolumn{1}{c}{} & \multicolumn{1}{c}{} & \multicolumn{1}{c}{$1R_{\rm d}$} & \multicolumn{1}{c}{$2R_{\rm d}$} & \multicolumn{1}{c}{$1R_{\rm d}$} & \multicolumn{1}{c}{$2R_{\rm d}$} & \multicolumn{1}{c}{phot} & \multicolumn{1}{c}{kin} & \multicolumn{1}{c}{kin}  \\ [5pt]
\hline
1 & 11.21 & 304.59 & 4.31 & 6.83 & 14.30 & 0.26 & 1.41 & 0.57 & 2.73 & 0.18 & 0.13 & 0.37 \\ [3.5pt]
2 & 11.16 & 343.62 & 3.24 & 6.77 & 12.35 & 0.18 & 0.81 & 0.24 & 1.07 & 0.28 & 0.12 & 0.21 \\ [3.5pt]
3 & 11.08 & 283.95 & 3.28 & 6.54 & 13.50 & 0.24 & 1.16 & 0.21 & 0.84 & 0.25 & 0.13 & 0.32 \\ [3.5pt]
4 & 10.89 & 258.21 & 5.31 & 5.98 & 14.29 & 0.31 & 1.19 & 0.57 & 1.92 & 0.25 & 0.23 & 0.37 \\ [3.5pt]
5 & 10.79 & 214.50 & 5.24 & 5.70 & 15.55 & 0.38 & 1.07 & 0.64 & 2.41 & 0.26 & 0.32 & 0.51 \\ [3.5pt]
6 & 10.78 & 235.98 & 5.57 & 5.68 & 15.47 & 0.19 & 0.99 & 0.36 & 1.23 & 0.34 & 0.17 & 0.28 \\ [3.5pt]
7 & 10.77 & 226.11 & 5.21 & 5.65 & 15.53 & 0.23 & 0.89 & 0.26 & 0.87 & 0.30 & 0.14 & 0.28 \\ [3.5pt]
8 & 10.57 & 194.36 & 2.53 & 5.06 & 13.06 & 0.22 & 0.90 & 0.26 & 0.92 & 0.20 & 0.19 & 0.32 \\ [3.5pt]
9 & 10.50 & 169.03 & 3.58 & 3.43 & 10.99 & 0.15 & 0.72 & 0.25 & 0.99 & 0.52 & 0.33 & 0.47 \\ [3.5pt]
10 & 10.49 & 181.28 & 4.46 & 4.83 & 11.39 & 0.30 & 0.84 & 0.28 & 1.21 & 0.07 & 0.24 & 0.43 \\ [3.5pt]
11 & 10.42 & 160.28 & 4.03 & 4.19 & 10.89 & 0.22 & 0.79 & 0.54 & 1.54 & 0.26 & 0.34 & 0.49 \\ [3.5pt]
12 & 10.40 & 159.31 & 5.12 & 4.40 & 12.12 & 0.21 & 0.75 & 0.27 & 0.92 & 0.38 & 0.30 & 0.43 \\ [3.5pt]
13 & 10.37 & 169.11 & 3.82 & 3.36 & 12.05 & 0.16 & 0.88 & 0.29 & 1.23 & 0.41 & 0.34 & 0.55 \\ [3.5pt]
14 & 10.36 & 155.43 & 3.25 & 2.75 & 9.21 & 0.11 & 0.71 & 0.20 & 1.00 & 0.48 & 0.37 & 0.59 \\ [3.5pt]
15 & 10.34 & 168.51 & 3.03 & 3.00 & 13.17 & 0.15 & 0.81 & 0.22 & 0.89 & 0.22 & 0.20 & 0.41 \\ [3.5pt]
16 & 10.28 & 148.47 & 4.60 & 3.66 & 11.90 & 0.17 & 0.80 & 0.28 & 1.07 & 0.48 & 0.41 & 0.54 \\ [3.5pt]
17 & 10.24 & 125.72 & 3.95 & 2.31 & 9.94 & 0.36 & 0.86 & 0.67 & 1.88 & 0.21 & 0.67 & 0.67 \\ [3.5pt]
18 & 10.19 & 125.94 & 5.90 & 3.99 & 12.83 & 0.32 & 1.06 & 0.49 & 1.63 & 0.32 & 0.50 & 0.58 \\ [3.5pt]
19 & 10.19 & 127.86 & 3.64 & 2.51 & 9.76 & 0.21 & 0.77 & 0.45 & 1.37 & 0.30 & 0.61 & 0.71 \\ [3.5pt]
20 & 10.16 & 152.94 & 3.42 & 3.35 & 8.01 & 0.20 & 0.89 & 0.25 & 1.11 & 0.30 & 0.29 & 0.44 \\ [3.5pt]
21 & 10.14 & 129.34 & 6.57 & 3.85 & 12.63 & 0.19 & 0.73 & 0.43 & 0.99 & 0.44 & 0.44 & 0.51 \\ [3.5pt]
22 & 10.08 & 132.25 & 4.70 & 3.65 & 10.92 & 0.19 & 0.68 & 0.28 & 0.84 & 0.38 & 0.42 & 0.51 \\ [3.5pt]
23 & 10.07 & 107.29 & 7.59 & 3.63 & 14.90 & 0.17 & 0.79 & 0.29 & 0.92 & 0.46 & 0.54 & 0.61 \\ [3.5pt]
24 & 10.06 & 110.89 & 5.36 & 3.60 & 12.15 & 0.26 & 0.83 & 0.35 & 1.12 & 0.43 & 0.51 & 0.59 \\ [3.5pt]
25 & 10.04 & 127.39 & 4.20 & 3.52 & 10.16 & 0.23 & 0.72 & 0.55 & 1.23 & 0.51 & 0.51 & 0.64 \\ [3.5pt]
26 & 10.01 & 131.04 & 3.96 & 3.47 & 10.81 & 0.25 & 0.85 & 0.30 & 0.85 & 0.28 & 0.33 & 0.48 \\ [3.5pt]
27 & 10.00 & 122.00 & 4.80 & 3.44 & 13.36 & 0.24 & 0.99 & 0.30 & 1.08 & 0.25 & 0.40 & 0.58 \\ [3.5pt]
\hline 
\hline 

\end{tabular*}

\raggedright
\tablenotetext{a}{The identification number of the NewHorizon galaxy.}
\tablenotetext{b}{The stellar mass inside three effective radii (log-scale) in the unit of solar mass.}
\tablenotetext{c}{The circular velocity of the galaxy in the unit of $\rm km/s$.}
\tablenotetext{d}{The effective radius in the unit of kpc.}
\tablenotetext{e}{The disk scale radius in the unit of kpc.}
\tablenotetext{f}{The radius that contains the 90\% of the total stellar mass in the unit of kpc.}
\tablenotetext{g}{The photometric vertical scale height of the thin disk, measured at 1$R_{\rm d}$ or 2$R_{\rm d}$ in the unit of kpc.}
\tablenotetext{h}{The photometric vertical scale height of the thick disk, measured at 1$R_{\rm d}$ or 2$R_{\rm d}$ in the unit of kpc.}
\tablenotetext{i}{The $r$-band luminosity fraction of the thick disk from the photometric or the kinematic decomposition, measured at $0.5\,-\,2R_{\rm d}$.}
\tablenotetext{j}{The mass fraction of the thick disk from the kinematic decomposition, measured at $0.5\,-\,2R_{\rm d}$.}

\end{table}

\begin{figure*}
\includegraphics[width=0.9\textwidth]{FigureA5.pdf}
\caption{
Similar to Figure~\ref{fig_chemical}-(b) in the main text but for the rest of the NH2 sample.
} 
\label{fig_appendixA5}
\end{figure*}

\begin{figure*}
\includegraphics[width=0.9\textwidth]{FigureA6.pdf}
\caption{
Similar to Figure~\ref{fig_chemical}-(c) in the main text but for the rest of the NH2 sample.
} 
\label{fig_appendixA6}
\end{figure*}

\begin{figure*}
\includegraphics[width=0.9\textwidth]{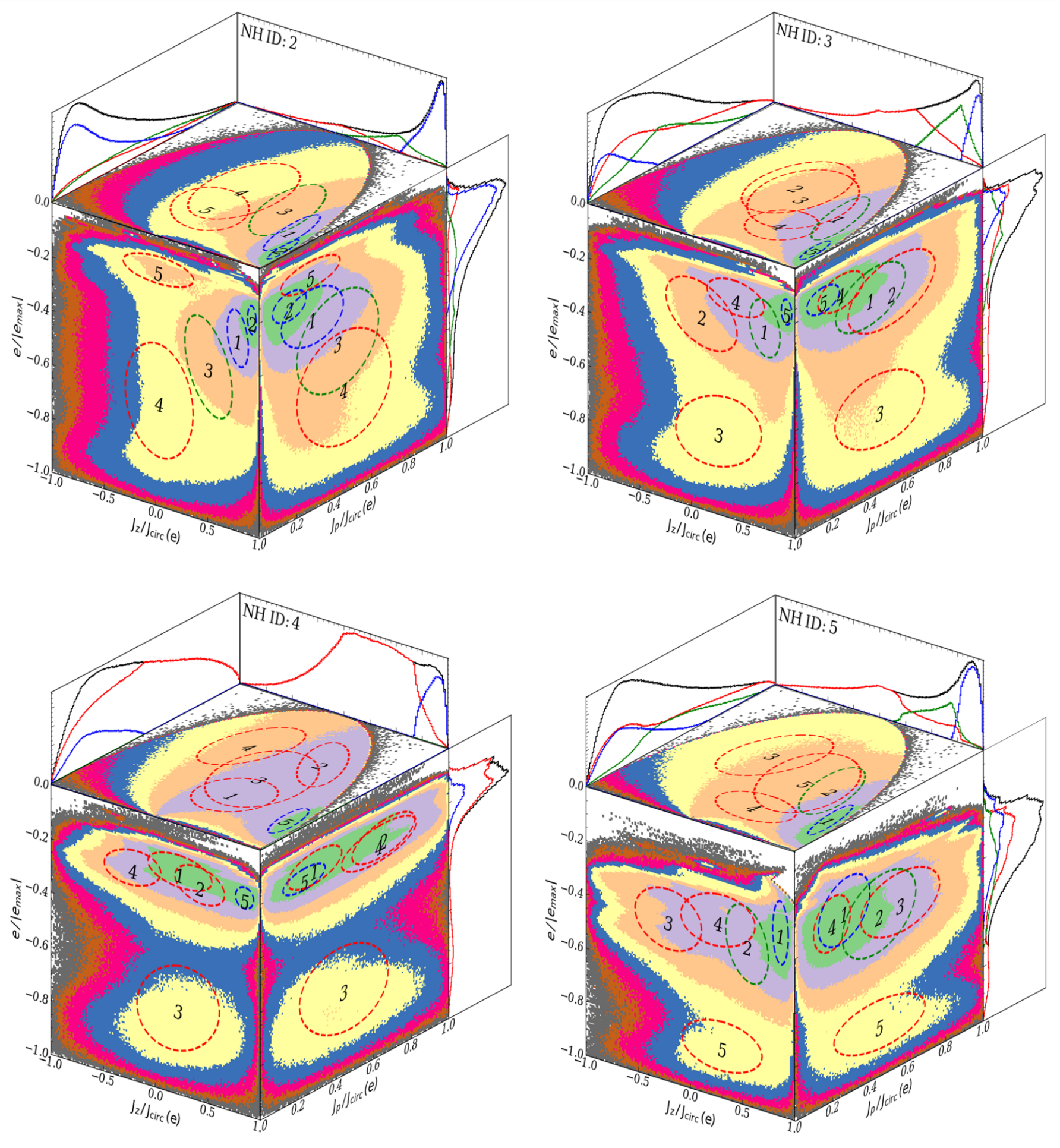}
\caption{
Same as Figure~\ref{fig_gmm} in the main text but for additional NH galaxies for reference. 
} 
\label{fig_appendixA7}
\end{figure*}

\begin{figure*}
\includegraphics[width=0.9\textwidth]{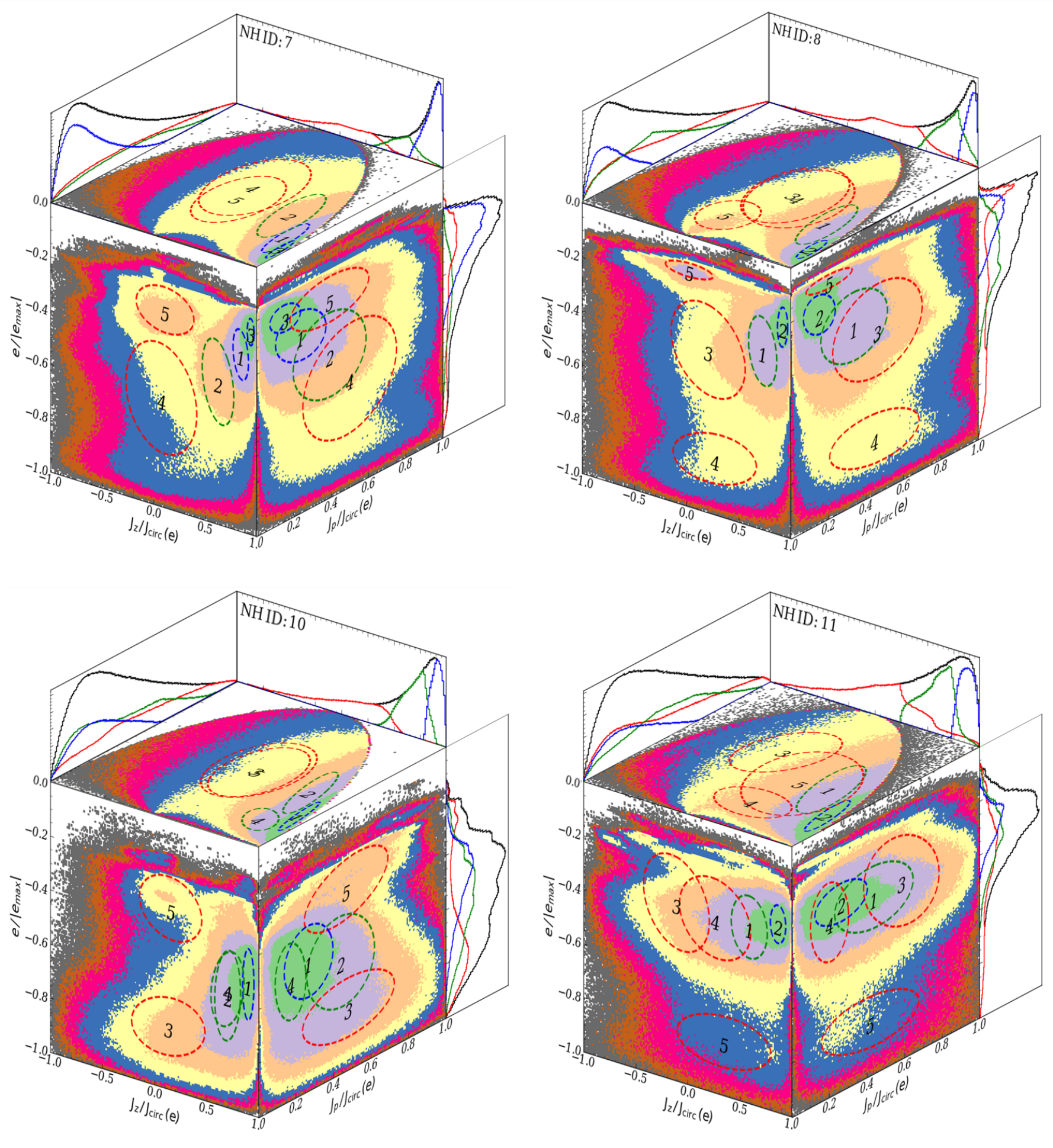}
\caption{
Same as Figure~\ref{fig_appendixA7} but for different galaxies. 
} 
\label{fig_appendixA8}
\end{figure*}

\begin{figure*}
\includegraphics[width=0.9\textwidth]{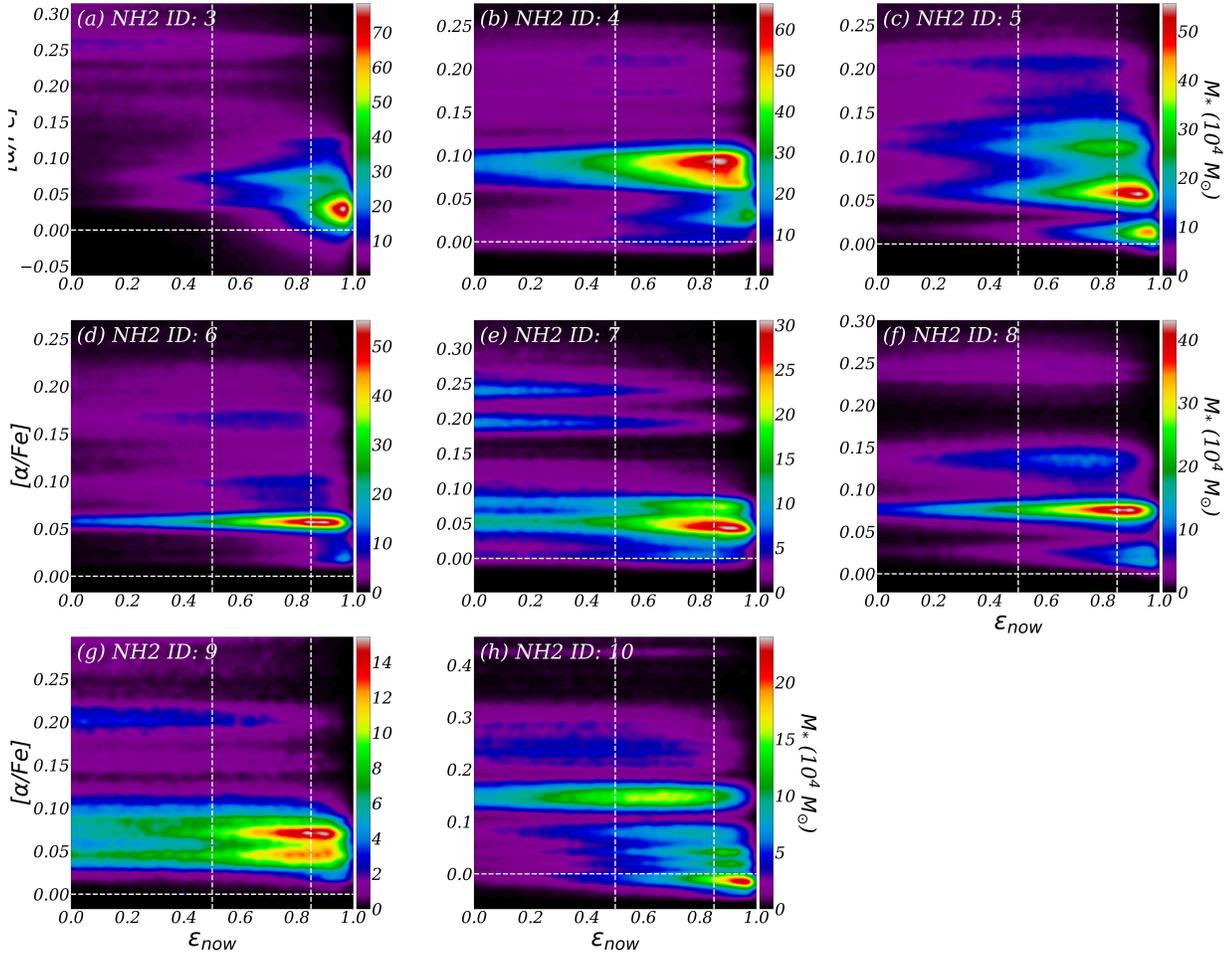}
\caption{
Same as Figure~\ref{fig_alpha_circ} in the main text but for the rest of the NH2 sample.
}
\label{fig_appendixA9}
\end{figure*}

\begin{figure*}
\includegraphics[width=0.9\textwidth]{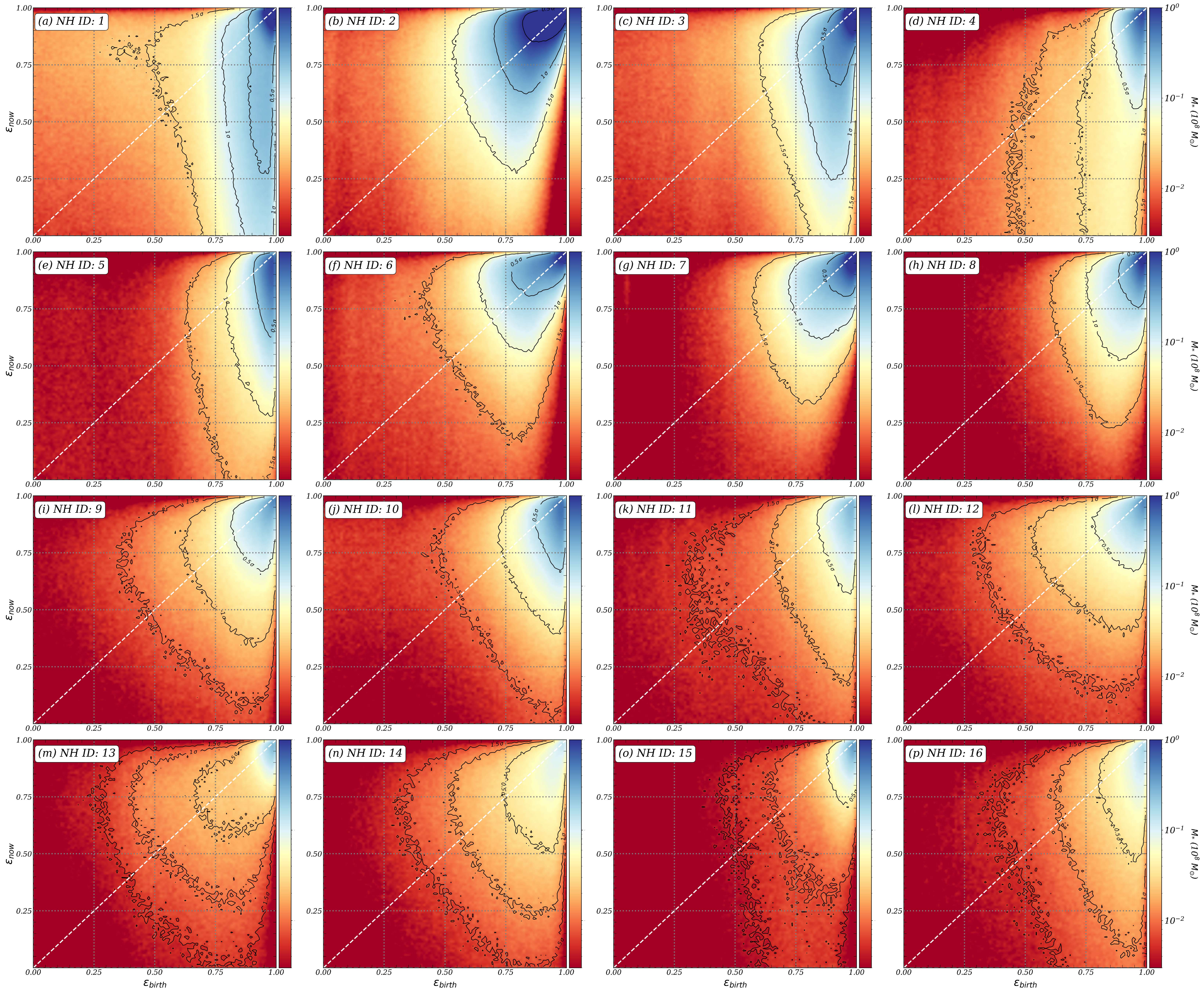}
\caption{
Same as Figure~\ref{fig_circ_birthnow} in the main text but for the 16 most massive NH galaxies.
} 
\label{fig_appendixA10}
\end{figure*}

\end{document}